\documentclass[letterpaper,11pt]{JHEP3}
\usepackage{bm}
\usepackage{amsmath}
\usepackage{epsfig}
\usepackage{psfrag}
\usepackage{cite}
\usepackage{verbatim}
\usepackage{yfonts}

\newcommand{\qn}{{\textswab{q}}}
\newcommand{\wn}{{\textswab{w}}}

\def\x{{\bm x}}
\def\k{{\bm k}}
\renewcommand{\Re}{\mathrm{Re}\,}
\renewcommand{\Im}{\mathrm{Im}\,}
\def\ofo{ { {}_2 \! F_1 }}
\def\A{{\scriptscriptstyle A}}
\def\B{{\scriptscriptstyle B}}
\def\C{{\scriptscriptstyle C}}
\def\D{{\scriptscriptstyle D}}

\title{Quasinormal modes and holography}

\author{ Pavel K.~Kovtun \\KITP,
University of California, Santa Barbara, CA 93106-4030, USA \\
E-mail: \email{kovtun@kitp.ucsb.edu}
}
\author{ Andrei O.~Starinets\\
Perimeter Institute for Theoretical Physics,
Waterloo, ON N2L 2Y5, Canada\\
E-mail: \email{starina@perimeterinstitute.ca}
}

\preprint{NSF-KITP-05-41}

\abstract{
Quasinormal frequencies of electromagnetic and gravitational
perturbations in asymptotically AdS spacetime
can be identified with poles of the corresponding
real-time Green's functions
in a holographically dual finite temperature field theory.
The quasinormal modes are defined for gauge-invariant
quantities which obey incoming-wave boundary condition
at the horizon and Dirichlet condition at the boundary.
As an application, we explicitly find poles
of retarded correlation functions
of $R$-symmetry currents and the energy-momentum tensor
in strongly coupled finite temperature
${\cal N}=4$ supersymmetric $SU(N_c)$ Yang-Mills theory
in the limit of large $N_c$.
}

\keywords{AdS/CFT correspondence, thermal field theory}

\normalsize

\begin{document}
\section{Introduction}
\label{intro_section}

While the statistical origin of black hole entropy remains a
subject of  active research, one may wonder if the celebrated analogy
\cite{BH}
between the laws of black hole mechanics and the laws of thermodynamics
can be generalized to non-equilibrium processes.
Holographic AdS/CFT correspondence
(\cite{Maldacena:1997re,Gubser:1998bc,Witten:1998qj}, 
see \cite{Aharony:1999ti} for a review)
provides a suitable arena for such a generalization.
AdS/CFT conjecture
asserts that string theories on certain asymptotically anti de Sitter
spacetimes are dual to quantum field theories in
lower dimension.
Since the low-energy limit of string theory is described by
the appropriate supergravity,
problems in general relativity can be mapped to
problems in the dual field theory.
According to the duality,
asymptotically AdS background spacetimes with event horizons
are interpreted as thermal states in dual field theories.
Correspondingly,
small perturbations of a black hole or a black brane background
are interpreted as small deviations from thermodynamic equilibrium
in a dual theory. This particular entry in the holographic dictionary can be
made  precise by considering
quasinormal spectra of asymptotically AdS spacetimes.

Quasinormal modes (see reviews \cite{Kokkotas:1999bd} and references therein)
are solutions to linearized equations obeyed by classical fluctuations
of a gravitational background subject to specific boundary conditions.
The choice of the boundary condition at the (future)
 horizon is dictated by the fact that
classically horizons do not emit radiation. Thus out of two local solutions
near the horizon typically representing waves incoming to the horizon and
outgoing from it,
one chooses the incoming waves only.
This choice of the boundary condition has profound consequences,
making the boundary value problem non-Hermitian, and
the associated eigenfrequencies complex.
This, however, is exactly what one expects in a holographically
dual theory, where
small deviations from thermal equilibrium are described by
dispersion relations which correspond to non-zero damping
\cite{Horowitz:1999jd}.
Mathematically, these dispersion relations appear
%(at least in the low-frequency limit)
as singularities of the retarded%
\footnote{
	Choosing outgoing waves at the horizon, one obtains
	advanced Green's functions in the dual theory.}
Green's functions in the complex frequency plane.
The connection between quasinormal spectrum of AdS black holes and
singularities of thermal correlators in dual quantum field theories
was first noted and explored for $2+1$ dimensional BTZ black holes in
\cite{Birmingham:2001pj}.
It was pointed out later \cite{Son:2002sd} that, even for higher-dimensional
systems, imposing Dirichlet boundary conditions for
scalar perturbations at asymptotic infinity ensures that
quasinormal frequencies coincide with the singularities
of the retarded Green's function in a holographically dual theory.

Quasinormal modes for electromagnetic and gravitational perturbations
are physically more interesting than those for scalars
because the corresponding fluctuations couple to
conserved symmetry currents in the dual quantum field theory.
However, the relation of these quasinormal modes to correlation functions
of the dual theory is not immediate: for example,
choosing Dirichlet boundary conditions for gauge-dependent
quantities such as metric perturbations would be rather unnatural.
Thus, we address the following question in this paper:
considering  computation of a quasinormal spectrum
as a purely general relativity problem
(independent of the holographic duality),
what variables and boundary conditions should one use
in order to ensure that the
resulting spectrum coincides with the poles of the correlators in the dual
quantum field theory?
(For related discussions, see
\cite{Moss:2001ga,Nunez:2003eq,Konoplya:2003dd}.)

For vector and gravitational fluctuations, a convenient approach similar to
the one used in cosmology \cite{bar} is to work with
gauge-invariant combinations of the fluctuations.
As an example, consider gravitational fluctuations
$h_{\mu\nu}$ of five-dimensional AdS-Schwarzschild background with
translationally invariant horizon.
According to the gauge theory/gravity duality the fluctuation couples
to the stress-energy tensor on the boundary \cite{Aharony:1999ti}, 
and thus we expect
the quasinormal spectrum of
$h_{\mu\nu}$ to be related to the poles of the retarded
two-point correlation function
$G_{\mu\nu,\alpha\beta}$ of the stress-energy tensor. As we
discuss in Section \ref{field_theory}, the two-point
function of the stress-energy tensor in the dual theory
is a sum of three independent
components
\begin{displaymath}
   G_{\mu\nu,\alpha\beta}(k) =
   S_{\mu\nu,\alpha\beta}\, G_1(k_0,\k^2) +
   Q_{\mu\nu,\alpha\beta}\, G_2(k_0,\k^2) +
   L_{\mu\nu,\alpha\beta}\, G_3(k_0,\k^2) \,,
\end{displaymath}
where $k$ is the four-momentum, and
$S_{\mu\nu,\alpha\beta}$, $Q_{\mu\nu,\alpha\beta}$,
$L_{\mu\nu,\alpha\beta}$ are the appropriate projectors
which provide three independent Lorentz index structures.
The correlation function is therefore completely determined by three scalar
functions $G_1$, $G_2$, $G_3$. On the gravity side, one can form
three gauge-invariant combinations of the components of $h_{\mu\nu}$
in such a way that each of them corresponds to one of the functions
$G_1$, $G_2$, $G_3$. Each of these gauge-invariant variables satisfies
a second-order ordinary differential equation
whose connection matrix essentially determines
$G_1$, $G_2$, $G_3$. 
Dirichlet boundary condition imposed on each of the
gauge-invariant variables ensures
that their quasinormal spectrum corresponds to the poles of
$G_1$, $G_2$, $G_3$.

The paper is organized as follows. In Section \ref{field_theory}
we discuss general Lorentz index structure of thermal
correlators of conserved currents and stress-energy tensor
in relativistic quantum field theories.
In Section \ref{sec:quasiholo}
we propose a way to identify
quasinormal frequencies of asymptotically AdS
spacetimes with poles of the corresponding
retarded Green's functions in the holographically dual
finite temperature field theory.
In Section \ref{sec:applications} we provide a detailed account
of scalar, electromagnetic and gravitational quasinormal spectra
for the five dimensional AdS-Schwarzschild background
with translationally invariant horizon.
Using the approach of gauge-invariant variables,
we explicitly show that one can define
quasinormal modes whose frequencies coincide with
singularities of retarded Green's functions
in the dual theory, when the latter are
computed using the standard AdS/CFT prescription.
In the low-energy limit we reproduce earlier results
\cite{Policastro:2001yc,Policastro:2002se,Policastro:2002tn}
on hydrodynamic properties of $3+1$ dimensional ${\cal N}{=}4$
supersymmetric $SU(N_c)$ Yang-Mills theory.
In the more general case, we numerically compute
the positions (in the complex frequency plane)
of the singularities of retarded
correlation functions of global $R$-symmetry currents
and energy-momentum tensor in strongly coupled ${\cal N}{=}4$
supersymmetric Yang-Mills theory in the limit of large $N_c$.
Some technical details appear in two appendices.

\section{Field theory correlators}

\label{field_theory}

We start by discussing Lorentz index structure of
retarded Green's functions of conserved currents
and energy-momentum tensor in relativistic
quantum field theories in infinite, flat $D$-dimensional
Minkowski space.
Translation and rotation invariance are assumed to be
unbroken symmetries of the theory.
We shall be interested in retarded Green's functions of
conserved symmetry currents,
\begin{equation}
    C_{\mu\nu}(x-y) =
    -i\,\theta(x^0{-}y^0) \langle [J_\mu(x),J_\nu(y)]\rangle
\end{equation}
as well as of stress-energy tensor,
\begin{equation}
    G_{\mu\nu,\alpha\beta}(x-y) =
    -i\,\theta(x^0{-}y^0) \langle [T_{\mu\nu}(x),T_{\alpha\beta}(y)]\rangle \ .
\end{equation}
The expectation value is taken in a translation-invariant state,
so that the expressions can be Fourier transformed:
\begin{equation}
    C_{\mu\nu}(x-y) = \int \frac{d^Dk}{(2\pi)^D}\;
    e^{ik(x-y)} C_{\mu\nu}(k)
\end{equation}
and similarly for $G_{\mu\nu,\alpha\beta}(x-y)$.
Here $k=(k_0,\k)$ is a $D$-dimensional momentum vector,
$kx=k_0 x^0 +\k{\bm x}$,
and the metric $\eta_{\mu\nu}$ is taken to be mostly plus.
Expectation values of all global conserved charges are
assumed to vanish in the equilibrium state,
in other words we consider systems without chemical potentials.
Then CPT invariance of the equilibrium state implies that
\begin{eqnarray}
   && C_{\mu\nu}(k) = C_{\nu\mu}(k) \ ,
   \label{eq:symm-C}\\
   && G_{\mu\nu,\alpha\beta}(k) = G_{\alpha\beta,\mu\nu}(k)\ .
   \label{eq:symm-G}
\end{eqnarray}
In addition, correlation functions of stress-energy tensor satisfy
\begin{equation}
    G_{\mu\nu,\alpha\beta}(k) =
    G_{\nu\mu,\alpha\beta}(k) =
    G_{\mu\nu,\beta\alpha}(k)
    \label{eq:symm-T}
\end{equation}
because of the symmetry of $T_{\mu\nu}(x)$.
Conservation of $J_\mu(x)$ and $T_{\mu\nu}(x)$ implies that
the correlation functions may be defined so that
they satisfy the following Ward identities%
\footnote{
    One may choose to define the correlation functions in such a way that
    local (in position space) counter-terms appear on the right-hand side
    of the Ward identities.
    The correlation functions defined in this way will differ from
    $C_{\mu\nu}(k)$ and $G_{\mu\nu,\alpha\beta}(k)$ by analytic functions
    of $k_0$ and $\k$.}
\begin{eqnarray}
   &&k^\mu C_{\mu\nu}(k) = 0\ ,
   \label{eq:WI-C}\\
   &&k^\mu G_{\mu\nu,\alpha\beta}(k) = 0\ .
   \label{eq:WI-G}
\end{eqnarray}
If in addition the theory possesses scale invariance,
then the correlation functions of stress-energy tensors
satisfy an extra Ward identity
\begin{equation}
   \eta^{\mu\nu} G_{\mu\nu,\alpha\beta}(k) = 0 \ .
\label{eq:WI-trace}
\end{equation}
Hermiticity of $J_\mu$ and $T_{\mu\nu}$ combined with
rotation invariance implies
\begin{eqnarray}
    && C_{\mu\nu}(k_0,\k) = C_{\mu\nu}(-k_0,\k)^*\,,
\label{eq:hermiticity-J}\\
    && G_{\mu\nu,\alpha\beta}(k_0,\k) = G_{\mu\nu,\alpha\beta}(-k_0,\k)^*\,.
\label{eq:hermiticity-T}
\end{eqnarray}

\subsection{Conserved currents}

In vacuum, the Ward identity (\ref{eq:WI-C})
implies that the current-current correlation function
$C_{\mu\nu}(k)$ is proportional to the
projector onto conserved vectors,
\begin{equation}
    P_{\mu\nu} = \eta_{\mu\nu} - \frac{k_\mu k_\nu}{k^2}\,,
\label{eq:Pmunu}
\end{equation}
where $k^2=-k_0^2+\k^2$.
All components of $C_{\mu\nu}(k)$ are thus determined
by a single scalar function,
\begin{equation}
    C_{\mu\nu}(k) = P_{\mu\nu}\, \Pi(k^2) \ .
\label{eq:C-Lorentz-inv}
\end{equation}
If the expectation value is taken in a state that has only
rotation symmetry (such as thermal equilibrium state
in canonical ensemble),
it is convenient to split the projector $P_{\mu\nu}$ into
transverse and longitudinal parts,
\begin{equation}
   P_{\mu\nu} = P_{\mu\nu}^T + P_{\mu\nu}^L\,,
\end{equation}
where $P_{\mu\nu}^T$ and $P_{\mu\nu}^L$
are mutually orthogonal
($P^T_{\alpha\mu}\eta^{\mu\nu}P^L_{\nu\beta}=0$)
projectors defined as
\begin{eqnarray}
   && P_{00}^T = 0\ , \ \ \
      P_{0i}^T = 0\ , \ \ \
      P_{ij}^T = \delta_{ij} - \frac{k_i k_j}{\k^2}\ ,
      \label{eq:PmunuT-def}\\
   && P_{\mu\nu}^L = P_{\mu\nu} - P_{\mu\nu}^T\ .
      \label{eq:PmunuL-def}
\end{eqnarray}
They satisfy $k^\mu P_{\mu\nu}^T = k^\mu P_{\mu\nu}^L = 0$,
and therefore any expression constructed out of
$P_{\mu\lambda}^T$, $P_{\mu\lambda}^L$ will automatically
satisfy current-conservation constraint.
Therefore in the rotation-invariant case the
current-current correlation function
is determined by two independent scalar functions,
\begin{equation}
   C_{\mu\nu}(k) = P_{\mu\nu}^T\, \Pi^T(k_0,\k^2) +
                   P_{\mu\nu}^L\, \Pi^L(k_0,\k^2)\ .
\label{eq:C-rotation-inv}
\end{equation}
When $\Pi^T=\Pi^L=\Pi$, this expression reduces to
the Lorentz-invariant form (\ref{eq:C-Lorentz-inv}).
Also, it is not difficult to see that due to rotation invariance
the $\k \to 0$ limits (with $k_0$ fixed)
of transverse and longitudinal self-energies coincide,
\begin{equation}
    \lim_{\k\to0}\Pi^T(k_0,\k^2) = \lim_{\k\to0}\Pi^L(k_0,\k^2)\,.
\label{eq:small-q-current}
\end{equation}
As an example, consider a four-dimensional
field theory at non-zero temperature.
Without loss of generality one can take the spatial momentum
oriented along the $x^3$ direction, so that $k_\mu=(-\omega,0,0,q)$,
with $k^2=-\omega^2+q^2$.
Then the components
of the current-current correlation function are
\begin{equation}
   C_{x^1 x^1}(k) = C_{x^2 x^2}(k) = \Pi^T(\omega,q)\ ,
\end{equation}
as well as
\begin{equation}
   C_{tt}(k){=} \frac{q^2}{\omega^2{-}q^2}\, \Pi^L(\omega,q),\ \
   C_{tx^3}(k){=} \frac{-\omega q}{\omega^2{-}q^2}\, \Pi^L(\omega,q),\ \
   C_{x^3 x^3}(k){=} \frac{\omega^2}{\omega^2{-}q^2}\, \Pi^L(\omega,q)\ .
     \label{eq:Czz}
\end{equation}
For a system in stable thermodynamic equilibrium at temperature $T$,
the low-energy ($\omega{\ll}T$, $q{\ll}T$) behavior of
$\Pi^T$ and $\Pi^L$ is universal and is described by
effective hydrodynamic theory
(see for example \cite{Forster}).
In the approximation of linearized hydrodynamics,
$\Pi^T(\omega,q)$ is non-singular as a function of $\omega$
because it does not couple to charge density fluctuations.
On the other hand, correlators which involve
conserved charge density must exhibit a hydrodynamic singularity
whose dispersion relation satisfies $\omega(q){\to}0$ as $q{\to}0$.
The longitudinal self-energy  $\Pi^L(\omega,q)$ has a simple pole at
$\omega{=}-i D_{\rm\scriptscriptstyle Q}\, q^2$,
where $D_{\rm\scriptscriptstyle Q}$ is the diffusion constant
of charge $Q$ associated with current $J_\mu(x)$.

\subsection{Stress-energy tensor}

In  vacuum, the two-point correlation function of
stress-energy tensor may be written as a sum of five terms
allowed by the symmetries (\ref{eq:symm-G}), (\ref{eq:symm-T}),
which are proportional to
$\eta_{\mu\nu}\eta_{\alpha\beta}$,
$(\eta_{\mu\alpha}\eta_{\nu\beta}+\eta_{\mu\beta}\eta_{\nu\alpha})$,
$(\eta_{\mu\nu}k_\alpha k_\beta + k_\mu k_\nu \eta_{\alpha\beta})$,
$(\eta_{\mu\alpha}k_\nu k_\beta + \eta_{\nu\alpha}k_\mu k_\beta
+ \eta_{\mu\beta}k_\nu k_\alpha + \eta_{\nu\beta}k_\mu k_\alpha)$, and
$k_\mu k_\nu k_\alpha k_\beta$.
Only two linear combinations of these terms are consistent with
the Ward identity (\ref{eq:WI-G});
they can be taken to be
$P_{\mu\nu}P_{\alpha\beta}$ and
$P_{\mu\alpha}P_{\nu\beta} + P_{\mu\beta}P_{\nu\alpha}$.
A convenient way to write $G_{\mu\nu,\alpha\beta}(k)$ is
\begin{equation}
    G_{\mu\nu,\alpha\beta}(k) =
       P_{\mu\nu}P_{\alpha\beta}\, G_B(k^2) +
       H_{\mu\nu,\alpha\beta}\, G_S(k^2) \ ,
\label{eq:G-Lorentz-inv}
\end{equation}
where
$$
    H_{\mu\nu,\alpha\beta} =
       \frac12\left(P_{\mu\alpha}P_{\nu\beta}+
       P_{\mu\beta}P_{\nu\alpha}\right)-
       \frac{1}{D{-}1}P_{\mu\nu}P_{\alpha\beta}
$$
is a projector onto conserved traceless symmetric tensors,
which is
constructed to satisfy $\eta^{\mu\nu}H_{\mu\nu,\alpha\beta}=0$.
As a result, a scale-invariant theory must have $G_B(k^2)=0$, and
the correlation function takes a simple form
\begin{equation}
    G_{\mu\nu,\alpha\beta}(k) = H_{\mu\nu,\alpha\beta}\, G_S(k^2)\ .
\label{eq:G-Lorentz-scale-inv}
\end{equation}
If the expectation value is taken in a state that
has only rotation symmetry (such as thermal equilibrium
state in the canonical ensemble), it is convenient to
split $H_{\mu\nu,\alpha\beta}$ into
mutually orthogonal projectors
constructed out of $P_{\mu\nu}^T$, $P_{\mu\nu}^L$.
One relevant combination is
\begin{equation}
  S_{\mu\nu,\alpha\beta} = \frac12 \left(
  P_{\mu\alpha}^T P_{\nu\beta}^L + P_{\mu\alpha}^L P_{\nu\beta}^T +
  P_{\mu\beta}^T P_{\nu\alpha}^L + P_{\mu\beta}^L P_{\nu\alpha}^T
  \right)\,.
\end{equation}
It satisfies $k^\mu S_{\mu\nu,\alpha\beta}=0$, and
also $\eta^{\mu\nu}S_{\mu\nu,\alpha\beta}=0$
because of the orthogonality of $P^T$ and $P^L$.
It is not difficult to find
another independent combination with the same properties,
\begin{equation}
  Q_{\mu\nu,\alpha\beta} = \frac{1}{D{-}1} \left(
  (D{-}2)P_{\mu\nu}^L P_{\alpha\beta}^L +
  \frac{1}{D{-}2}P_{\mu\nu}^T P_{\alpha\beta}^T -
  (P_{\mu\nu}^T P_{\alpha\beta}^L + P_{\mu\nu}^L P_{\alpha\beta}^T)
  \right) \ .
\end{equation}
The projectors $S_{\mu\nu,\alpha\beta}$ and $Q_{\mu\nu,\alpha\beta}$
square to themselves,
and are orthogonal to each other
($S_{\mu\nu,\alpha\beta}\,\eta^{\alpha\lambda}\, \eta^{\beta\rho}
  Q_{\lambda\rho,\sigma\tau} = 0$).
Therefore the projector $H_{\mu\nu,\alpha\beta}$ can be split as
$$H_{\mu\nu,\alpha\beta} = S_{\mu\nu,\alpha\beta} +
  Q_{\mu\nu,\alpha\beta} + L_{\mu\nu,\alpha\beta}\ ,$$
where $L_{\mu\nu,\alpha\beta}\equiv H_{\mu\nu,\alpha\beta}-
S_{\mu\nu,\alpha\beta}-Q_{\mu\nu,\alpha\beta}$
is orthogonal to both $S_{\mu\nu,\alpha\beta}$ and $Q_{\mu\nu,\alpha\beta}$.
Thus the correlation function of energy-momentum tensor
in a scale-invariant theory can be written as a sum over
three independent index structures,
\begin{equation}
   G_{\mu\nu,\alpha\beta}(k) =
   S_{\mu\nu,\alpha\beta}\, G_1(k_0,\k^2) +
   Q_{\mu\nu,\alpha\beta}\, G_2(k_0,\k^2) +
   L_{\mu\nu,\alpha\beta}\, G_3(k_0,\k^2) \ .
\label{eq:G-scale-rotation-inv}
\end{equation}
It is not difficult to show that rotation invariance implies
that $\k \to 0$ limits (with $k_0$ fixed) of the three
independent scalar functions must coincide,
\begin{equation}
    \lim_{\k\to0}G_1(k_0,\k^2) =
    \lim_{\k\to0}G_2(k_0,\k^2) = 
    \lim_{\k\to0}G_3(k_0,\k^2)\,.
\label{eq:small-q-stress}
\end{equation}
In a scale non-invariant theory, two extra scalar functions
are needed to specify $G_{\mu\nu,\alpha\beta}(k)$.
They multiply two independent linear combinations of
$P_{\mu\nu}^T P_{\alpha\beta}^T$,
$P_{\mu\nu}^L P_{\alpha\beta}^L$, and
$(P_{\mu\nu}^T P_{\alpha\beta}^L + P_{\mu\nu}^L P_{\alpha\beta}^T)$;
one possible choice is
{\setlength\arraycolsep{2pt}
\begin{eqnarray}
   G_{\mu\nu,\alpha\beta}(k)
   &=&
   \left( P_{\mu\nu}^T P_{\alpha\beta}^T
   +\frac12 (P_{\mu\nu}^T P_{\alpha\beta}^L + P_{\mu\nu}^L P_{\alpha\beta}^T)
   \right) C_T(k_0,\k^2) \nonumber\\
   &+&
   \left( P_{\mu\nu}^L P_{\alpha\beta}^L
   +\frac12 (P_{\mu\nu}^T P_{\alpha\beta}^L + P_{\mu\nu}^L P_{\alpha\beta}^T)
   \right) C_L(k_0,\k^2) \nonumber\\
   &+&
   S_{\mu\nu,\alpha\beta}\, G_1(k_0,\k^2) +
   Q_{\mu\nu,\alpha\beta}\, G_2(k_0,\k^2) +
   L_{\mu\nu,\alpha\beta}\, G_3(k_0,\k^2) \ .
\end{eqnarray}
}%
When $C_T{=}C_L{=}G_B$ and $G_1{=}G_2{=}G_3{=}G_S$,
this expression reduces to the Lorentz-invariant form
(\ref{eq:G-Lorentz-inv}).

As an example, consider a four-dimensional
field theory at non-zero temperature.
Choosing momentum to be $k_\mu =(-\omega,0,0,q)$ as above,
one finds the following components of the correlation function.
The correlations of transverse momentum density
are determined by $G_1(\omega,q)$,
\begin{eqnarray}
  &&G_{tx^1,tx^1}(k)
    = \frac12 \frac{q^2}{\omega^2{-}q^2} G_1(\omega,q) ,\label{eq:Gtxtx}\\
  &&G_{tx^1,x^1x^3}(k)
    =-\frac12 \frac{\omega q}{\omega^2-q^2} G_1(\omega,q) ,\\
  &&G_{x^1x^3,x^1x^3}(k)
    =\frac12 \frac{\omega^2}{\omega^2{-}q^2} G_1(\omega,q) \label{eq:Gxzxz} .
\end{eqnarray}
The correlations of longitudinal momentum density,
energy density, and diagonal stress are determined by $G_2(\omega,q)$,
$C_L(\omega,q)$, $C_T(\omega,q)$.
For example,
\begin{eqnarray}
  &&G_{tt,tt}(k)= \frac13 \frac{q^4}{(\omega^2{-}q^2)^2}
    \Big[2G_2(\omega,q) + 3C_L(\omega,q) \Big] \ ,
  \label{eq:Gtttt}\\
  &&G_{tt,tx^3}(k)= -\frac13 \frac{\omega q^3}{(\omega^2{-}q^2)^2}
    \Big[2G_2(\omega,q) + 3C_L(\omega,q) \Big] \ ,
  \label{eq:Gtttz}\\
%  &&G_{tx^3,tx^3}(k)= \frac13 \frac{\omega^2 q^2}{(\omega^2{-}q^2)^2}
%    \Big[2G_2(\omega,q) + 3C_L(\omega,q)\Big]\ ,
%  \label{eq:Gtztz} \\
  &&G_{tt,x^1 x^1} = \frac16 \frac{q^2}{(q^2{-}\omega^2)}
    \Big[2G_2(\omega,q) - 3C_L(\omega,q) - 3C_T(\omega,q)\Big] \ .
  \label{eq:Gttxx}
\end{eqnarray}
The correlations of transverse stress
are determined by $G_3(\omega,q)$,
\begin{equation}
    G_{x^1 x^2,x^1 x^2}(k) = \frac12 G_3(\omega,q)\,.
\label{eq:Gxyxy}
\end{equation}
For a system in stable thermodynamic equilibrium at temperature $T$,
the low-energy ($\omega{\ll}T$, $q{\ll}T$) behavior of
$G_{\mu\nu,\alpha\beta}$ is universal and is described by
effective hydrodynamic theory
(see for example \cite{Forster}).
In the approximation of linearized hydrodynamics,
$G_3(\omega,q)$ is non-singular as a function of $\omega$
because it does not couple to energy density
or momentum density fluctuations.
On the other hand, correlation functions
which involve conserved densities exhibit
hydrodynamic singularities whose dispersion relations satisfy
$\omega(q){\to}0$ as $q{\to}0$.
Function $G_1(\omega,q)$ has a simple pole at $\omega{=}-i\gamma_\eta q^2$,
where $\gamma_\eta$ is the damping constant of the shear mode,
proportional to shear viscosity.
In a conformal theory (when $C_L=C_T=0$),
function $G_2(\omega,q)$ has simple poles at
$\omega=\pm v_s q -i\Gamma_{\!s} q^2$,
where $v_s$ is the speed of sound, and
$\Gamma_{\!s}$ is the damping constant of the sound mode,
also proportional to shear viscosity.

\section{Quasinormal spectrum and holographic duality}
\label{sec:quasiholo}
\subsection{Thermal correlation functions}

We will be interested in small fluctuations of a black $p$-brane,
\begin{equation}
    ds^2 = a(r) \left( -f(r) dt^2 +
    \sum_{i=1}^p (d x^i)^2\right)  + b(r) dr^2 \,.
\label{metric}
\end{equation}
Metrics of this form arise as a result of dimensional reduction
of higher dimensional supergravity backgrounds.
In addition, these backgrounds have non-zero values
of various ``matter'' fields which we collectively denote
by $\phi^{(0)}$ suppressing all indices.
Holographically dual theory
is defined on the boundary ($r{\to}\infty$)
of (\ref{metric}), which is a flat
$p{+}1$ dimensional Minkowski space.

Translation invariance on the boundary implies that
all fluctuating fields can be taken to be proportional to
$e^{-i\omega t+i\bm{qx}}$;
thus linearized fluctuations $\delta g_{\mu\nu}$, $\delta \phi$ of
the background will obey a system of
second-order linear ordinary differential equations.
Generically, the system will be redundant, reflecting the
gauge freedom (such as linearized diffeomorphisms)
enjoyed by the fluctuation fields. Instead of fixing a particular gauge,
we consider gauge-invariant combinations
of the fluctuation fields. 
Let $Z_k$ be these gauge-invariant variables
which are constructed as linear combinations
of fluctuating fields and their derivatives, excluding
$r$-derivatives.
Variables $Z_k$ will obey a system of coupled second order linear 
ordinary differential equations (ODEs)
which can in principle be diagonalized. Let $Z(r)$ be
such a gauge-invariant variable satisfying
a second-order ODE.
A local solution of the ODE near the horizon will generally be a superposition
of incoming and outgoing waves. Since classically the horizon does not
radiate, we choose the incoming wave boundary condition there.

The solution obeying the incoming wave boundary condition at the horizon
can be written in the basis of two local solutions at the boundary%
\footnote{
	By local solutions we mean the solutions obtained as
	power series around the corresponding singular point
	($r{=}\infty$ in this case) of the differential equation.
	See for example \cite{ince} for a general discussion.
}
as
\begin{equation}
Z(r) = {\cal A}\,  \varphi_1(r) + {\cal B}\,  \varphi_2(r) \,,
\label{general}
\end{equation}
where ${\cal A}$, ${\cal B}$ are the connection
coefficients of the corresponding ODE. Coefficients  ${\cal A}$, ${\cal B}$
typically depend on the parameters (such as frequency and momentum)
which enter the differential equation for $Z(r)$.
Near the boundary, the solution (\ref{general}) becomes
\begin{equation}
     Z(r) =
     {\cal A}\, r^{- \Delta_-}\; \left( 1 + \cdots\right) +
     {\cal B}\, r^{-\Delta_+} \;  \left( 1 + \cdots\right)\,,
\end{equation}
where $\Delta_+$, $\Delta_-$ are exponents of the ODE at $r=\infty$,
and ellipses denote higher powers of $r$.
We consider the situation when the exponents are not equal,
$\Delta_+>\Delta_-$, and $\Delta_+$ can be taken positive.%
\footnote{
	In AdS/CFT correspondence, $\Delta_+$ is equal to
	the conformal dimension of the operator that couples
	to bulk fields contained in $Z$.
	See also \cite{Klebanov:1999tb} for exceptional cases.
}

The action of the system expanded to quadratic order in
fluctuations $\delta g_{\mu\nu}$,
 $\delta \phi$ can be rewritten in terms of the gauge-invariant variables.
On shell, the part of the action quadratic in fluctuations
will reduce to the boundary term
\begin{equation}
   S^{(2)} \sim \lim_{r\to\infty} \int\!d\omega d^p\!q \,
   F(\omega,{\bm q})\, Z'(r)\, Z(r) + \mbox{contact terms}\,,
\label{action_gen}
\end{equation}
where $F(\omega,{\bm q})$ depends on the details of the action,
and ``contact terms'' do not contain $Z'(r)$.

In holographic AdS/CFT duality, fluctuation $\delta \phi$ couples to a
particular operator ${\cal O}$ of the dual theory at the boundary.
Applying the Lorentzian AdS/CFT
prescription \cite{Son:2002sd, Herzog:2002pc}
to the action (\ref{action_gen})
to compute the retarded correlator, and remembering that $Z(r)$ is a
functional of  $\delta \phi$, we find
\begin{equation}
\langle {\cal O} {\cal O}\rangle_R
 \sim \frac{\cal B}{\cal A} +\mbox{contact terms}\,.
\label{holo_corr}
\end{equation}
The poles of the retarded correlator correspond to zeros of the
connection coefficient ${\cal A}$.
On the other hand, setting ${\cal A}=0$ in Eq.~(\ref{general})
corresponds to a particular choice of boundary conditions
for the fluctuation $Z(r)$. From the general relativity point of view,
this choice determines the quasinormal spectrum of $Z(r)$.
In other words, equation ${\cal A}=0$ defines
quasinormal spectrum for gauge-invariant perturbations
which has the interpretation of the poles
of retarded correlators in a holographically dual theory.%
\footnote{
\label{fn:exceptional-poles}
One should add at once that % :-)
in addition to zeros of ${\cal A}$, singularities of the
correlator (\ref{holo_corr}) may also come from singularities of
${\cal B}$. However, singularities of ${\cal B}$ are completely
determined by the singularities of the {\it local} Frobenius solution
$\varphi_1$ considered as a function of parameter(s) of the
differential equation. Indeed, a general theorem \cite{arnold}
guarantees smoothness of a solution
of a differential equation with respect to a parameter, if the equation and
the boundary conditions depend smoothly on the parameter. Thus the
solution $Z$ is smooth, and singularities of  ${\cal B}$ are
destined to cancel the singularities of the coefficients
of the series expansion in the local
Frobenius solution  $\varphi_1$ with respect to a parameter.
Therefore, singularities of ${\cal B}$ are essentially determined
by the recursion relations
 for  the coefficients of the series expansion which defines $\varphi_1$.
Let us illustrate this point using hypergeometric equation as an
example. On the interval $z\in [0,1]$, the solution $\ofo
(a,b;c;z)$ defined by its recursion relations at $z=0$ is related
to two local solutions at $z=1$ by
\begin{eqnarray}
\ofo \left( a,b; c;z\right) &=& {\cal A}\; \ofo \left( a,b; a+b-c+1;1-z\right)
\nonumber \\&+&
 {\cal B}\; (1-z)^{c-a-b} \, \ofo \left( c-a, c-b; c-a-b+1;1-z\right)\,,
\label{eqnop}
\end{eqnarray}
where the connection matrix coefficients are given by
\begin{displaymath}
 {\cal A} = \frac{\Gamma (c) \Gamma (c-a-b)}{\Gamma (c-a)\Gamma (c-b)}\,,
\;\;\;\;\;\; \;\;\;\;  {\cal B}
 = \frac{\Gamma (c) \Gamma (a+b-c)}{\Gamma (a)\Gamma (b)}\,.
\end{displaymath}
To be in agreement with  the scenario set by Eq.~(\ref{general}),
let us further assume that $c-a-b>0$.  The poles of the correlator
(\ref{holo_corr}) come from the poles of $\Gamma (c-a)$ and
 $\Gamma (c-b)$ (also corresponding to the Dirichlet condition
${\cal A}=0$), {\it and} from the poles of $\Gamma (a+b-c)$.
The latter are determined by the local solution near $z=1$,
\begin{eqnarray}
\ofo \left( a,b; a+b-c+1;1-z\right) &=&
 1 + \frac{a b}{1+a+b-c} (1-z)\nonumber \\
&+&
\frac{a b (a+1) (b+1)}{2 (1+a+b-c)(2+a+b-c)} (1-z)^2 +\cdots ,
\nonumber
\end{eqnarray}
the coefficients of which have poles at $a+b-c=-n$, where $n$ is a
positive integer. These are precisely the poles of the correlator
(\ref{holo_corr}) coming from  ${\cal B}$. By setting $a+b-c = -n
+\epsilon$ and taking the limit $\epsilon \rightarrow 0$, one can
show that the right hand side of Eq.~(\ref{eqnop}) is in fact a
smooth function of the parameters $a$, $b$, $c$.
}
This argument is similar to the one given in
\cite{Danielsson:1999zt, Son:2002sd} in the case of a scalar
fluctuation, the difference being the use of gauge-invariant
variables in the present discussion.

\subsection{Black brane fluctuations}
\label{classif}

According to the dictionary of the AdS/CFT correspondence
\cite{Aharony:1999ti},
global symmetry currents of the dual field theory
have as their sources boundary values of the gauge field
$A_\mu$ on the higher-dimensional background (\ref{metric}).
Similarly, stress-energy tensor of the dual theory
is sourced by gravitational fluctuations $h_{\mu\nu}$
of the black brane.
We shall be interested in quasinormal spectra of these fluctuations.
We take the fluctuations to be of the form
$A_\mu(r)e^{-i\omega t+i q z}$,
$h_{\mu\nu}(r)e^{-i\omega t+i q z}$ where $z=x^p$.
The fluctuations  can be classified according to their transformation
properties under the ``remaining'' world-volume symmetry group
$O(p{-}1)$ acting on $x^1,..,x^{p-1}$.

Let us restrict ourselves to the simplest case when the considered
fluctuations do not couple to fluctuations of other background fields.
Components $A_{t}$, $A_{z}$ do not transform under $O(p{-}1)$, while
components  $A_{\alpha}$, $\alpha = x^1,..,x^{p-1}$ transform as vectors.
One can therefore distinguish two symmetry channels
for electromagnetic fluctuations
\begin{subequations}
\label{class_vect}
\begin{eqnarray}
&\,&
\mbox{Spin 0 \; (diffusive channel):} \hspace{0.5cm}
 A_t,\, A_z,\, A_r  \\
&\,&
\mbox{Spin 1 \; (transverse  channel):}  \hspace{0.5cm}
  A_\alpha\,. 
\end{eqnarray}
\end{subequations}
A similar classification can be adopted for metric fluctuations.
Components $h_{tt}$, $h_{tz}$, $h_{zz}$, $h_{rr}$, $h_{tr}$, and $h_{zr}$
do not transform under $O(p{-}1)$,
components $h_{t\alpha}$, $h_{z\alpha}$, and $h_{r\alpha}$
transform as vectors, 
while $h_{\alpha\beta}$ transform as rank-2 tensors.
The tensor representation is reducible, for a symmetric $h_{\alpha\beta}$
can be decomposed into the trace part
$\delta_{\alpha\beta} h/(p{-}1)$,
where $h=\sum_{\alpha} h_{\alpha\alpha}$ (a singlet)
and the symmetric traceless part
$h_{\alpha\beta}-\delta_{\alpha\beta} h/(p{-}1)$.
Thus we have three symmetry channels for gravitational fluctuations:%
\footnote{
	The name ``scalar'' for spin-$2$ fluctuations reflects the fact
	that the corresponding wave equation coincides with that of the
	minimally coupled massless scalar \cite{Kovtun:2004de}.
	The names ``shear'' and ``sound'' (as well as ``diffusive'' above)
	reflect physical interpretation of the lowest quasinormal frequency
	(for a given symmetry channel) in the dual field theory.
	This will be seen explicitly in the next section.
}
\begin{subequations}
\label{class_grav}
\begin{eqnarray}
&& \mbox{Spin 0 (sound channel):} \hspace{0.5cm}
 h_{tt},\, h_{tz},\, h_{zz},\, h,\, h_{rr},\, h_{tr},\, h_{zr} \\
&& \mbox{Spin 1 (shear channel):}  \hspace{0.5cm}
  h_{t\alpha},\,  h_{z\alpha},\, h_{r\alpha} 
\\
&& \mbox{Spin 2 (scalar channel):}  \hspace{0.5cm}
 h_{\alpha\beta}-\delta_{\alpha\beta} h/(p{-}1)\,.
\end{eqnarray}
\end{subequations}
The $O(p{-}1)$ symmetry guarantees that equations for fluctuations belonging
to different symmetry channels decouple.
Classification presented here mirrors the classification of the correlators
in Section \ref{field_theory}: diffusive and transverse channels of the
$U(1)$ fluctuation correspond to functions $\Pi^L$ and  $\Pi^T$
of the current-current correlator.
When the dual theory possesses conformal invariance,
shear, sound and scalar channels of the
gravitational fluctuation are related to functions $G_1$, $G_2$,
$G_3$ of the stress-energy tensor correlator.

\subsection{Gauge-invariant variables}

We now define gauge-invariant variables corresponding to classes
(\ref{class_vect}), (\ref{class_grav}). 
Combinations of the gauge field components invariant under
the transformation
$A_{\mu}\rightarrow A_{\mu} - \partial_{\mu} \lambda$ include
components of the electric field
\begin{eqnarray}
  \mbox{Diffusive channel:} \hspace{0.5cm}
E_z &=& q \, A_t + \omega \, A_z\,,\\
  \mbox{Transverse channel:} \hspace{0.5cm} E_\alpha &=& \omega A_\alpha\,.
\end{eqnarray}
Perturbations  $h_{\mu\nu}$ transform under infinitesimal
diffeomorphisms as
$h_{\mu\nu}{\to}h_{\mu\nu} -
 \nabla_{\!\mu}\,\xi_{\nu} -
 \nabla_{\!\nu}\,\xi_{\mu}$,
where $\xi_\mu = \xi_\mu (r) e^{-i\omega t+iqz}$ are gauge functions,
and covariant derivatives are taken with respect to
the background metric (\ref{metric}).
One may define the following gauge-invariant combinations
linear in perturbations:
\begin{eqnarray}
    &\,& \mbox{Shear channel:}  \hspace{0.2cm}
    Z_1 = q H_{tx^1} + \omega H_{zx^1}\label{l2}
\label{l1}\\
    &\,& \mbox{Sound channel:} \hspace{0.2cm}
    Z_2 = q^2 f H_{tt} + 2 \omega q H_{tz} +\omega^2 H_{zz} +
    q^2 f \left( 1 + \frac{a f'}{a' f} - \frac{\omega^2}{q^2 f}\right)H\, \\
    &\,& \mbox{Scalar channel:}  \hspace{0.2cm}
    Z_3 = H_{x^1x^2}\,,
\label{l3}
\end{eqnarray}
where $H_{tt}= h_{tt}/af$, $H_{tz}=h_{tz}/a$, $H_{ij}= h_{ij}/a$ ($i,j\neq t$),
$H= h/(p{-}1)a$.
From Einstein equations obeyed by the fluctuations, one obtains
three independent second-order ODEs satisfied by $Z_1$, $Z_2$, and $Z_3$.

Quasinormal modes are defined as solutions to the second-order
differential equations satisfied by the gauge-invariant variables
$E_z$, $E_{\alpha}$, $Z_1$, $Z_2$, $Z_3$ obeying incoming wave
boundary condition at the horizon and Dirichlet condition at
the boundary.
According to the above discussion, 
the spectra of complex eigenfrequencies
obtained in solving the boundary value problem for fluctuations
$E_z$, $E_{\alpha}$, $Z_1$, $Z_2$, $Z_3$ coincide with poles of the
 functions $\Pi^L$, $\Pi^T$, $G_1$, $G_2$, $G_3$, respectively.

\section{Thermal correlators in strongly coupled ${\cal N}=4$ SYM}
\label{sec:applications}

We now apply the approach outlined in Section \ref{sec:quasiholo}
to the near-horizon limit of the non-extremal gravitational background of
type IIB low energy string theory describing
$N_c$ parallel black three-branes.
The background is given by the metric
\begin{equation}
  ds^2 = \frac{r^2}{R^2}\left( -f(r) dt^2 + dx^2 + dy^2 + dz^2 \right) +
         \frac{R^2}{r^2 f(r)} dr^2 + R^2 d\Omega_5^2\,,
\label{eq:D3-metric}
\end{equation}
where $R$ is a constant, which depends on the number of $D3$ branes,
$R\propto N_c^{1/4}$, and $f(r)=1-r_0^4/r^4$.
The parameter of non-extremality $r_0$ specifies the location of the horizon,
whose Hawking temperature is $T=r_0/\pi R^2$.
Upon dimensional reduction on $S^5$, gravitational perturbations
will effectively propagate on the five-dimensional part
of the background (\ref{eq:D3-metric}).
Introducing new coordinate $u=r_0^2/r^2$, the metric can be written as
\begin{equation}
   ds^2 =
    \frac{(\pi T R)^2}u
    \left( -f(u) dt^2 + dx^2 + dy^2 +dz^2 \right)
    +\frac{R^2}{4 u^2 f(u)} du^2 \,,
\label{near_horizon}
\end{equation}
where $f(u)=1-u^2$.
In these coordinates, the horizon is located at $u{=}1$,
and the boundary is at $u{=}0$.
In addition, the background is specified by
the value of the self-dual five form field
\begin{equation}
\label{non_extremal_5form}
%    F_5 = - \frac{4 (\pi T)^3 R^2}{u^{3/2}} (1 + * )\,
%    d t\wedge d x \wedge d y \wedge d z  \wedge d r\,,
    F_5 = \frac{2 (\pi T R)^4}{u^3} (1 + * )\,
    d t\wedge d x \wedge d y \wedge d z  \wedge d u\,,
\end{equation}
with all other fields vanishing.
The dual quantum field theory is ${\cal N}=4$ $SU(N_c)$
supersymmetric Yang-Mills theory
in $3+1$ dimensional Minkowski space at large $N_c$ and
large 't~Hooft coupling \cite{Maldacena:1997re}.
The field theory is taken in  a thermal equilibrium state
at a temperature equal to the Hawking temperature of the background.
Real-time thermal correlators of the conserved $R$-symmetry currents
and stress-energy tensor in this theory were considered in the
AdS/CFT approach in
\cite{Policastro:2002se, Policastro:2002tn, Nunez:2003eq}.
Here we show that
reformulating the problem in terms of gauge-invariant variables
allows one to compute functions $\Pi^L$, $\Pi^T$, $G_1$, $G_2$,
$G_3$ directly by solving the second-order ODEs associated with
each of them. In particular, quasinormal spectra of the
gauge-invariant fluctuations determine poles of these
functions in complex frequency plane.%
\footnote{
	The problem of additional singularities
	discussed in footnote \ref{fn:exceptional-poles}
	does not arise here.}

\subsection{$R$-current correlators}
\label{rcc}

Correlators of $R$-currents in  strongly coupled ${\cal N}=4$ SYM
at zero temperature were computed in
\cite{Freedman:1998tz, Chalmers:1998xr}
by using the AdS/CFT correspondence.
Similar approach can be taken at non-zero temperature \cite{Policastro:2002se}.
On the gravity side of AdS/CFT, one considers an effective $U(1)$ field%
\footnote{
	The $R$-charges of the theory $Q^a=\int\!d^3x\,j_0^a(x)$
	generate global $SU(4)$ symmetry group, with $a=1\dots15$.
	In an equilibrium state without chemical potentials
	for the $R$-charges, the correlation function of $R$ currents
	$j_\mu^a$ has the form
	$C_{\mu\nu}^{ab}(x) = \delta^{ab} C_{\mu\nu}(x)$.
	The expressions of this section refer to $C_{\mu\nu}(x)$.
}
in the five-dimensional asymptotically AdS part of the background
(\ref{near_horizon}).
This five-dimensional Maxwell field is essentially a graviphoton
of the dimensional reduction of (\ref{eq:D3-metric}) on $S^5$.

According to the discussion in Section \ref{field_theory},
thermal current-current correlators
are determined by two independent scalar functions,
$\Pi^T (\omega,q)$ and $\Pi^L (\omega,q)$. Correspondingly, we expect the
dual five-dimensional Maxwell system to reduce to
two independent equations
for gauge-invariant variables whose quasinormal frequencies
determine the poles of the correlators. 
One of the quasinormal frequencies should be purely imaginary (at least
in the regime $q/T \ll 1$),
reflecting diffusive relaxation of large-scale charge density fluctuations
around thermal equilibrium state in the dual field theory.

Maxwell's equations for the $U(1)$ field are simply
$\partial_\A(\sqrt{-g}\,g^{\A\C} g^{\B\D} F_{\C\D}){=}0$,
where $F_{\C\D}=\partial_\C A_\D - \partial_\D A_\C$,
capital Latin indices run over $t,x,y,z,u$, and
components of the five dimensional metric $g_{\A\B}$
are given by (\ref{near_horizon}).
Translation invariance for the $t,x,y,z$ coordinates
implies that vector potential can be Fourier transformed,
\begin{equation}
    A_\C(u,t,\x) =\int \frac{d^4k}{(2\pi)^4}
                  e^{ik_0 t + i\k\x} A_\C(u,k)\,.
\end{equation}
Choosing $k=(-\omega,0,0,q)$,
one can derive the following equations satisfied by
 transverse and longitudinal electric fields:
\begin{subequations}
\label{eq:electric-fields}
\begin{eqnarray}
    && E_{\alpha}'' + \frac{f'}{f} E_{\alpha}' +
       \frac{\wn^2-\qn^2 f}{u f^2} E_{\alpha} =0\,, 
       \; \; \; \alpha = x,y\,,
\label{eqE_x}\\
    && E_z '' + \frac{\wn^2 f'}{f (\wn^2 - \qn^2 f)} E_z' +
    \frac{\wn^2 - \qn^2 f}{u f^2} E_z = 0\,,
\label{eqE_z}
\end{eqnarray}
\end{subequations}
where $E_{\alpha}\equiv\wn A_{\alpha}$, $E_z\equiv\qn A_t + \wn A_z$,
dimensionless parameters $\wn$ and $\qn$ are defined as
\begin{equation}
     \wn = \frac{\omega}{2\pi T}\,, \qquad \qn = \frac{q}{2\pi T}\,,
\end{equation}
and prime denotes the derivative with respect to $u$.

Gauge-gravity duality implies
that all information about two-point $R$-current correlation functions
in the dual ${\cal N}{=}4$ SYM theory
(in the large $N_c$ and large 't Hooft coupling limit)
is contained in the
solutions to the differential equations (\ref{eq:electric-fields}).
For both equations,
the singularity at $u=1$ (the horizon) has
exponents $\pm i\wn/2$ corresponding to the outgoing/incoming waves.
To compute the retarded correlators, one has to impose the
incoming wave boundary condition at the horizon \cite{Son:2002sd}
thus choosing
$-i\wn/2$ as the correct exponent.
At the boundary ($u=0$) the exponents for both equations are $0$ and $1$,
and thus solutions to equations (\ref{eq:electric-fields})
which satisfy incoming-wave boundary condition at the horizon
behave near $u=0$ as
\begin{subequations}
\label{eq:boundary-E}
\begin{eqnarray}
    E_{\alpha} (u)&=&  {\cal A}_{(\alpha)}(\wn,\qn)+\cdots  +
    {\cal B}_{(\alpha)} (\wn,\qn)\,
    u +\cdots\,,\\
    E_z (u) &=&  {\cal A}_{(z)}(\wn,\qn) +
\cdots  + {\cal B}_{(z)} (\wn,\qn)
 \, u +\cdots\,.
\end{eqnarray}
\end{subequations}
The boundary action of the Maxwell system in the gauge
$A_u=0$ is%
\footnote{
	Normalization of the five-dimensional action
	$S=1/4g_B^2 \int \sqrt{-g}\, F_{\A\B} F^{\A\B}$
	is fixed by $g_B^2=16\pi^2R/N_c^2$ \cite{Freedman:1998tz}.
}
\begin{equation}
    S = \lim_{u\to0} \frac{N_c^2 T^2}{16} \int\frac{d\omega dq}{(2\pi)^2}
        \left[ A_t'(u,k) A_t(u,-k) -
        f(u)\, {\bm A}'(u,k) {\bm A}(u,-k)\right]
\end{equation}
Using Maxwell's equations, the action can be written
in terms of gauge-invariant variables as
\begin{align}
    S = \lim_{u\to0} & \frac{N_c^2 T^2}{16} \int\frac{d\omega dq}{(2\pi)^2}
        \Big[ \frac{f(u)}{\qn^2 f(u) -\wn^2} E_z'(u,k) E_z(u,-k) \nonumber\\
      & -\frac{f(u)}{\wn^2}
        \left( E_x'(u,k) E_x(u,-k) + E_y'(u,k) E_y(u,-k)\right)\Big]
        +\mbox{contact terms} \,,
\end{align}
where ``contact terms'' do not contain derivatives
of the electric fields.
In order to find the correlation functions,
one has to express the derivatives of the fields
in terms of the boundary values of the fields
$A_\mu^0(k){\equiv}A_\mu(u{\to}0,k)$,
$E_\mu^0(k){\equiv}E_\mu(u{\to}0,k)$
using the solutions (\ref{eq:boundary-E});
applying Lorentzian AdS/CFT prescription \cite{Son:2002sd},
one finds%
\footnote{
	See \cite{Son:2002sd} for the definition of
	functional derivative in this context.
}
\begin{equation}
  C_{\alpha\alpha}(\omega,q)  =
  \frac{\delta^2 S}{\delta A_{\alpha}^0 (k)\delta
  A_{\alpha}^0(-k)} =  \frac{ \wn^2
  \delta^2 S}{\delta E_{\alpha}^0 (k)\delta
  E_{\alpha}^0(-k)} =
  - \frac{N_c^2 T^2 {\cal B}_{(\alpha)}}{8 {\cal A}_{(\alpha)}}\,.
\label{eq:transverse-correlators}
\end{equation}
Similarly,
\begin{subequations}
\label{eq:longitudinal-correlators}
\begin{eqnarray}
  C_{tt}(\omega,q) &=&
  \frac{N_c^2 T^2 \qn^2 {\cal B}_{(z)}}{8 (\qn^2-\wn^2)
  {\cal A}_{(z)}}\,,
\label{ttc}\\
  C_{tz}(\omega,q) &=& C_{zt}(\omega,q)=
  \frac{N_c^2 T^2\wn\qn\,{\cal B}_{(z)}}{8 (\qn^2-\wn^2)\, {\cal A}_{(z)}}\,,
\label{tzc}\\
  C_{zz}(\omega,q) &=& \frac{N_c^2 T^2\wn^2\,{\cal B}_{(z)}}{8 (\qn^2-\wn^2)
  {\cal A}_{(z)}}\,.
\label{zzc}
\end{eqnarray}
\end{subequations}
Comparing these expressions with the general result (\ref{eq:Czz}),
one finds
\begin{equation}
   \Pi^T (\omega, q) = 
   - \frac{N^2_c T^2\, {\cal B}_{(\alpha)}(\omega,q)}
          {8\,{\cal A}_{(\alpha)}(\omega,q)}\,,
    \qquad
   \Pi^L (\omega, q) = 
   - \frac{N^2_c T^2\, {\cal B}_{(z)}(\omega,q)}
          {8\,{\cal A}_{(z)}(\omega,q)}\,.
\end{equation}
Thus the correlation functions are completely determined by the ratios
of the connection coefficients
of differential equations (\ref{eq:electric-fields}).
In particular,
poles of the correlators correspond to zeros of
the coefficients ${\cal A}_{(\alpha)}(\wn,\qn)$
and ${\cal A}_{(z)}(\wn,\qn)$.
To find the zeros, we impose Dirichlet boundary conditions
on electric fields at $u=0$
 for the solutions to equations (\ref{eq:electric-fields})
which satisfy the incoming wave conditions at the horizon.
Physically, the horizon acts as a perfectly absorbing surface,
while the boundary acts as a perfect conductor.

In order to
determine the self-energies $\Pi^T(\omega,q)$, $\Pi^L(\omega,q)$,
one needs to know the solution to equations (\ref{eq:electric-fields}).
Analytic solution is unknown, except for a special case of $\qn=0$,
when the Dirichlet boundary value problem
can be reformulated as a problem of solving a transcendental
algebraic continued fraction equation
\cite{Starinets:2002br, Nunez:2003eq}.
For $\qn=0$,  gauge-invariant variables
$E_i$, $i=x,y,z$ obey the same equation%
\footnote{
	As they should, in accord with rotation invariance
	in the dual field theory, see Eq. (\ref{eq:small-q-current}).
}
\begin{equation}
    E_i '' + \frac{f'}{f} E_i' + \frac{\wn^2}{u f^2} E_i = 0\,.
\label{eqE_zq}
\end{equation}
The solution to the Dirichlet boundary value problem
for Eq.~(\ref{eqE_zq}) can be found exactly, and is given by Heun
polynomials \cite{Nunez:2003eq}. The quasinormal spectrum is
\begin{equation}
    \qn = 0\,,\quad\quad  \wn = n (1 - i)\,, \quad n=0,1,2,...\,.
\label{poles_w}
\end{equation}
For $\qn\neq 0$, quasinormal spectra of perturbations $E_\alpha$
and $E_z$ (and, correspondingly, the poles of $\Pi^T (\omega,q)$
and $\Pi^L (\omega,q)$) differ from each other,
and can be found numerically, as explained in Appendix~\ref{app:numerics}.
A typical arrangement of quasinormal frequencies
is shown in Fig.~\ref{fig:complexplane-E-transverse}
(the poles of  $\Pi^L (\omega,q)$
were found numerically in \cite{Nunez:2003eq}).
\begin{FIGURE}[t]
{
\begin{minipage}[t]{.48\textwidth}
\begin{center}
  \psfrag{x}{$\scriptstyle \Re \wn $}
  \psfrag{y}{$\scriptstyle \Im \wn $}
  \epsfig{file=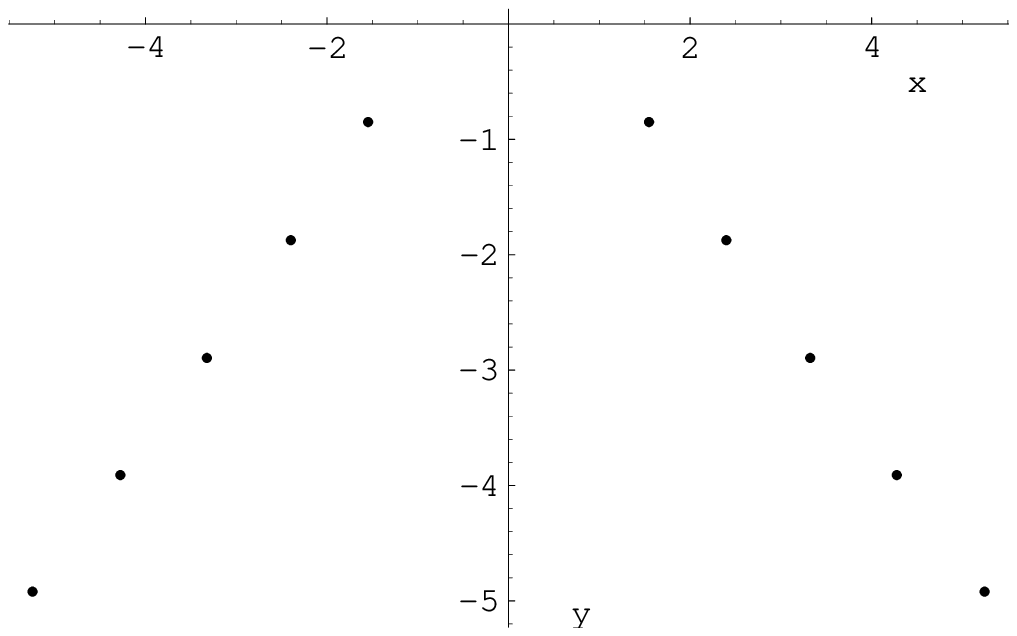,scale=0.7}
\end{center}
\end{minipage}
\begin{minipage}[t]{.48\textwidth}
\begin{center}
  \psfrag{x}{$\scriptstyle \Re \wn $}
  \psfrag{y}{$\scriptstyle \Im \wn $}
  \epsfig{file=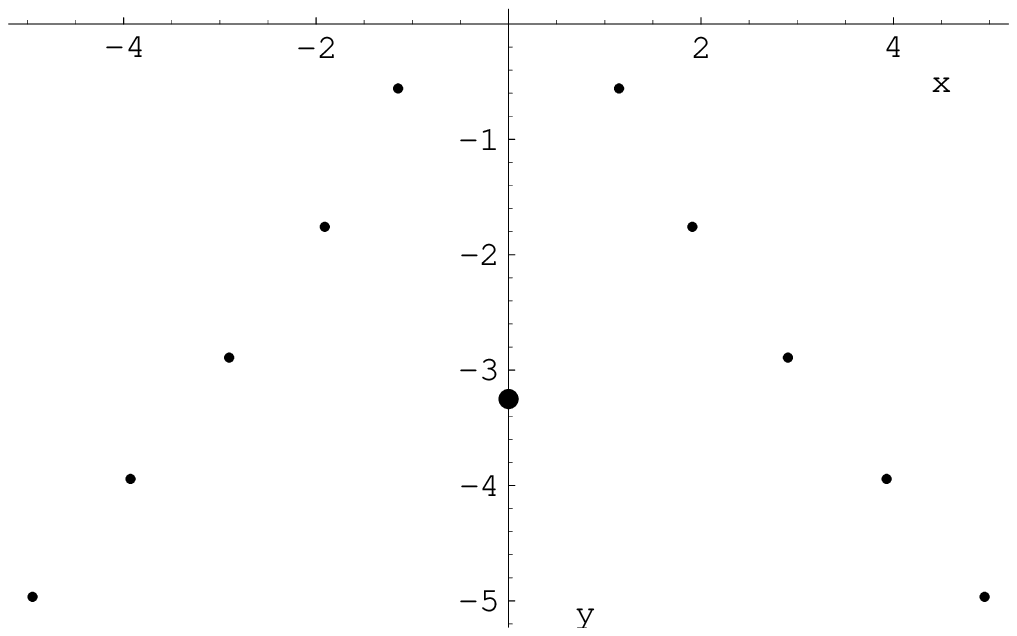,scale=0.7}
\end{center}
\end{minipage}
  \caption
    {%
	Quasinormal spectrum of electric field fluctuations
	in the plane of complex $\wn\equiv\omega/2\pi T$,
	shown for spatial momentum $\qn\equiv q/2\pi T = 1$.
	Quasinormal frequencies on the left are defined by equation
	(\ref{eqE_x}) for the transverse electric field, and coincide
	with the poles of $\Pi^T(\omega,q)$, as explained in the text.
	Quasinormal frequencies on the right
	are defined by equation (\ref{eqE_z}) for the
	longitudinal electric field, and coincide with the poles of
	$\Pi^L(\omega,q)$.
	As $\qn$ decreases, all poles stay at a finite distance
	from the real axis, except for the one marked with a large dot. 
        This pole is purely imaginary and approaches the origin
	in the limit $\qn\to0$.
	The presence of this special quasinormal frequency is a manifestation
	of the diffusive relaxation of $R$-charge density fluctuations
	in the dual ${\cal N}{=}4$ SYM theory.
    }
\label{fig:complexplane-E-transverse}
}
\end{FIGURE}%
Quasinormal frequencies are located symmetrically
with respect to the imaginary axis, as is expected
from the singularities of the corresponding correlation function
in the dual field theory, see equation (\ref{eq:hermiticity-J}).

\subsection*{Hydrodynamic approximation}
In the hydrodynamic limit
one can find analytic solutions to
equations (\ref{eq:electric-fields}) as a series in
$\wn\ll 1$, $\qn\ll 1$. Assuming first that $\wn$, $\qn$ are
of the same order, we get
\begin{subequations}
\label{eq:hydro-solutions-E}
\begin{eqnarray}
    E_{\alpha} (u)&=& C_{\alpha}\, f(u)^{-{i \wn}/{2}}\, \left[ 1+
    i \wn\,  \log{\frac{1+u}{2}} + O (\wn^2,\qn^2)\right]\,,
%\label{oo1}
\\
    E_{z} (u)&=& C_z\, f(u)^{-{i \wn}/{2}}\, \left[ 1 +
    i \wn \, \log{\frac{1+u}{2}} + \frac{i\qn^2}{\wn}\, (1-u)
    + O (\wn^2,\qn^2)\right]\,,
\label{eq:hydro-solution-Ez}
\end{eqnarray}
\end{subequations}
where $C_{\alpha}$, $C_z$ are normalization constants.
Imposing Dirichlet boundary conditions at $u=0$, one finds that
the equation $E_{\alpha} (0)=0$ has no solution compatible with the
assumption  $\wn\ll 1$. Holographic interpretation of this fact is that
the function  $\Pi^T(\omega,q)$ has no singularities
in the hydrodynamic regime.
The condition $E_{z} (0)=0$ leads to
$ \wn = -i \qn^2 + O(\qn^3)$.
This is the lowest hydrodynamic
quasinormal frequency of the
Dirichlet boundary value problem for $E_z$.
Accordingly, in the hydrodynamic regime
function $\Pi^L(\omega,q)$ has a pole at%
\footnote{
\label{fn:frequency-scaling}
  One may question the validity of this result
  since it implies that $\wn \sim \qn^2$, whereas the solution
  (\ref{eq:hydro-solution-Ez}) was obtained under the assumption
  that $\wn$ and $\qn$
  are of the same order. To check our result, we introduce a new
  parameter $\mu = \wn/\qn \sim \qn$. Solving Eq.~(\ref{eqE_z})
  perturbatively in $\mu \ll 1$, $\qn \ll 1$, where  $\mu$ and $\qn$
  are of the same order, we find
  $E_{z} (u)= C_z\, f^{-i \mu \qn/2}\, \left[ 1 +
  i \qn\, (1-u)/\mu + O(\mu)\right]$.
  The condition  $E_{z} (0)=0$ again gives the dispersion relation
  $\wn=-i\qn^2$.
}
\begin{equation}
     \omega = -i D_{\rm\scriptscriptstyle R} q^2\,,
\label{eq:diffusion-dispersion}
\end{equation}
where $D_{\rm\scriptscriptstyle R}=1/2\pi T$.
Physically, it corresponds to the
R-charge diffusion with the diffusion constant
$D_{\rm\scriptscriptstyle R}$.
By comparing the solutions (\ref{eq:hydro-solutions-E})
 to the definition of connection coefficients
(\ref{eq:boundary-E}), one finds
${\cal A}_{(\alpha)}=1$,
${\cal B}_{(\alpha)}=i\wn$, ${\cal A}_{(z)}=1+i\qn^2/\wn$,
${\cal B}_{(z)}=i\wn-i\qn^2/\wn$.
Substituting these connection coefficients
into AdS/CFT results
(\ref{eq:transverse-correlators}) and (\ref{eq:longitudinal-correlators}),
one reproduces the $R$-current correlators in the low-frequency
approximation found earlier in \cite{Policastro:2002se}.

\subsection{Stress-energy tensor correlators}

Correlation functions of stress-energy tensor in
the finite temperature  ${\cal N}=4$ SYM at large $N_c$ and strong coupling
were studied in
\cite{Policastro:2002se,Policastro:2002tn,Nunez:2003eq}.
Here we use the gauge-invariant variables approach to
identify the correct associated boundary value problem,
and to obtain new results for
the correlators in the sound wave channel.

To compute correlators of the stress-energy tensor in AdS/CFT correspondence,
one considers metric fluctuations $h_{\A\B}$ of the
supergravity background.
To linear order in $h_{\A\B}$, the
Einstein equations are
\begin{equation}
 {\cal R}_{\A\B}^{(1)} = - \frac{4}{R^2} h_{\A\B}\,,
\label{first_order_es}
\end{equation}
where ${\cal R}_{\A\B}^{(1)}$ is the linearized Ricci tensor
evaluated in the background (\ref{near_horizon}).
Translation invariance for the $t,x,y,z$ coordinates
implies that metric perturbations can be Fourier transformed,
and classified according to their transformations with respect to the
rotation group $O(2)$, as discussed in Section \ref{classif}.
As we shall see shortly, quasinormal spectra of the
three gauge-invariant variables (\ref{l1}) -- (\ref{l3})
appear correspondingly
as the poles of the three functions $G_1$, $G_2$, $G_3$ in
Eq.~(\ref{eq:G-scale-rotation-inv}) which determines
the two-point correlation function of stress-energy tensor
in a scale-invariant theory.

\subsubsection{Scalar channel}
According to the discussion in Section \ref{sec:quasiholo},
the equation satisfied by the component $h_{xy}$
of the perturbed metric decouples from the rest of
Einstein equations.
The gauge-invariant function $Z_3 = H_{xy}= h_y^x$ satisfies the
equation for a minimally coupled massless scalar in the background
(\ref{near_horizon}),
\begin{equation}
   Z_3'' - \frac{1+u^2}{uf} Z_3' + 
   \frac{\wn^2 -\qn^2 f}{uf^2} Z_3=0\,. 
\label{scalar_eq}
\end{equation}
The exponents of this equation near $u{=}0$ are $0$ and $2$,
therefore asymptotic behavior of $Z_3$ near the boundary is
\begin{equation}
    Z_3(u) = {\cal A}_{(3)} \left( 1 +\cdots \right) + 
          {\cal B}_{(3)} u^2 + \cdots\,,
\label{eq:boundary-Z3}
\end{equation}
where ellipses denote higher powers of $u$.
The relevant part of the boundary gravitational action
(Eq.(6.9) of \cite{Policastro:2002se}) can be written using
our notations as
\begin{equation}
   S = -\frac{\pi^2 N^2_c T^4}{8}\, \lim_{u\to 0}
   \int \frac{d \omega d q}{(2\pi)^2}\;
   \frac{f(u)}{u}\, Z_3'(u,k) Z_3(u,-k) \,.
\end{equation}
Proceeding as in Section {\ref{rcc}}, we obtain
the correlator%
\footnote{
	Terms analytic in $\wn$ and $\qn$ are ignored,
	even if they are divergent as $u{\to}0$.
}
\begin{equation}
     G_{xy,xy} = - \frac{\pi^2 N_c^2 T^4 {\cal B}_{(3)}}{2 {\cal A}_{(3)}}\,.
\label{eq:scalar-correlator}
\end{equation}
Comparing this expression to the general result (\ref{eq:Gxyxy}),
we find
\begin{equation}
    G_3(\omega,q) = - \frac{\pi^2 N_c^2 T^4\, {\cal B}_{(3)}(\omega,q)}
                      {{\cal A}_{(3)}(\omega,q)}\,.
\end{equation}
Turning now to the connection between quasinormal spectrum and
AdS/CFT correlators, we see that the condition ${\cal A}_{(3)}(\wn,\qn)=0$
determines the singularities of $G_3$.
From the definition of connection coefficients (\ref{eq:boundary-Z3})
it is evident that the condition ${\cal A}_{(3)}(\wn,\qn)=0$
is equivalent to imposing the
Dirichlet boundary condition on fluctuations, $Z_3(u{=}0)=0$.

The quasinormal spectrum of fluctuations obeying incoming wave
boundary condition at the horizon and Dirichlet condition
$Z_3(u{=}0)=0$ at the boundary was numerically computed in
\cite{Starinets:2002br, Nunez:2003eq}.
The spectrum $\wn_n$ is discrete, (presumably) infinite, and almost
equidistant. For $\qn=0$ its asymptotics for higher modes
is well approximated by a simple formula%
\footnote{
	For an analytic approach to the asymptotic behavior
	(\ref{asymptotics}) see \cite{schiappa}.
}
\cite{Starinets:2002br}
\begin{equation}
   \omega_n = 2\pi T n\, ( \pm 1 - i )\,, \;\;\;\;\;\; n\rightarrow
   \infty\,. \label{asymptotics}
\end{equation}
A typical arrangement of quasinormal frequencies
is shown in Fig.~\ref{fig:complexplane-scalar}.
\begin{FIGURE}[t]
{
  \parbox[c]{\textwidth}
  {
  \begin{center}
  \psfrag{x}{$\scriptstyle \Re \wn$}
  \psfrag{y}{$\scriptstyle \Im \wn$}
  \includegraphics[width=2.8in]{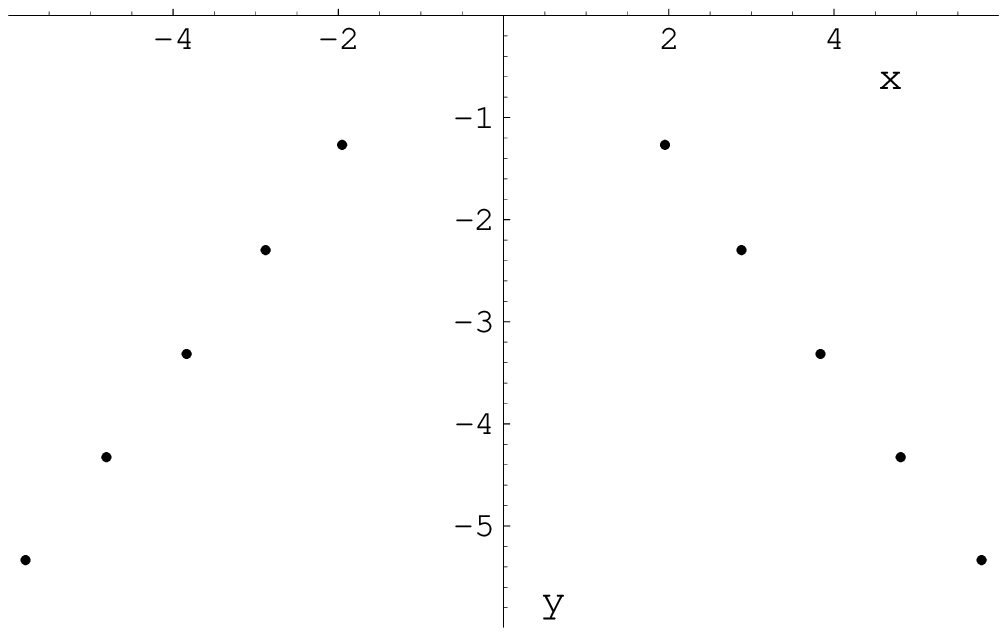}
  \caption
    {%
	Quasinormal spectrum of gravitational fluctuations
	in the scalar channel, shown in the plane of complex
	$\wn\equiv\omega/2\pi T$, for spatial momentum $\qn \equiv q/2\pi T=1$.
	The quasinormal frequencies coincide with poles of
	$G_3(\omega,q)$, as explained in the text.
	As $\qn\to 0$, all poles stay at a finite distance
	from the real axis, as one expects from the absence of
	hydrodynamic singularities in $G_3(\omega,q)$.
    }
  \end{center}
  }
\label{fig:complexplane-scalar}
}
\end{FIGURE}%
Quasinormal frequencies are located symmetrically
with respect to the imaginary axis, as is expected
from the singularities of the corresponding correlation function
in the dual field theory, see Eq.~(\ref{eq:hermiticity-T}).

\subsubsection*{Hydrodynamic approximation}

In the hydrodynamic limit $\wn\ll1$, $\qn \ll 1$,
analytic solution to Eq.~(\ref{scalar_eq}) can be found,
\begin{equation}
  Z_3(u) = C_3 \,f(u)^{-i\wn/2} \left[1 + O(\wn^2,\qn^2) \right]\,,
\label{eq:hydro-solution-Z3}
\end{equation}
where $C_3$ is a normalization constant.
By comparing this solution
with the definition of connection coefficients (\ref{eq:boundary-Z3}),
one finds
${\cal A}_{(3)} = 1 + O(\wn^2)$,
${\cal B}_{(3)} = {i\wn}/{2} + O(\wn^2, \qn^2, \wn\qn)$.
Substituting these connection coefficients into
the AdS/CFT result (\ref{eq:scalar-correlator}),
one reproduces the correlator of transverse components
of stress tensor in the low-frequency approximation
found earlier in \cite{Policastro:2002se}.
The equation ${\cal A}_{(3)}(\wn,\qn)=0$
does not have solutions compatible with the condition $\wn \ll 1$.
This is consistent with our expectations in Section \ref{field_theory}
that $G_3$ has no hydrodynamic singularities.

\subsubsection{Shear channel}

According to the discussion in  Section \ref{sec:quasiholo},
equations satisfied by the components
$h_{tx}$, $h_{zx}$, and $h_{ux}$
of the perturbed metric form a closed set.
In the radial gauge $h_{u\A}{=}0$, they read
\begin{subequations}
\begin{eqnarray}
    H_{zx}' &=& - \frac{\wn}{\qn f } H_{tx}'\,, \\
    H_{tx}'' &=& \frac{1}{u} H_{tx}' + \frac{\wn \qn}{u f} H_{zx} +
                 \frac{\qn^2}{u f} H_{tx}\,,
\end{eqnarray}
\end{subequations}
where $H_{tx} = u h_{tx}/(\pi T R)^2$, $H_{zx}=u h_{zx}/(\pi T R)^2$.
Using these equations, one finds that
$Z_1(u)\equiv \qn H_{tx}(u) + \wn H_{zx}(u)$ satisfies
the following second-order ODE
\begin{equation}
    Z_1'' + \frac{(\wn^2 - \qn^2 f)f - u\wn^2 f'}{uf(\qn^2 f -\wn^2)} Z_1'
    + \frac{\wn^2 - \qn^2 f}{u f^2} Z_1 = 0\,.
\label{okno}
\end{equation}
The exponents of Eq.~(\ref{okno}) at $u=0$ are $0$ and $2$,
and thus asymptotic behavior of $Z_1$ near the boundary is
\begin{equation}
   Z_1(u)={\cal A}_{(1)}(1+\dots) + {\cal B}_{(1)} u^2 +\dots\,,
\end{equation}
where ellipses denote higher powers of $u$.
The relevant part of the boundary gravitational action is given by
Eq.~(6.19) of \cite{Policastro:2002se}. Expressed in terms
of the gauge-invariant  variable $Z_1$, the action is
\begin{equation}
    S = -\frac{\pi^2 N^2_c T^4}{8}\, \lim_{u\rightarrow 0}
         \int \frac{d \omega d q}{(2\pi)^2}\;
         \frac{f(u)}{u (\wn^2 - \qn^2 f(u))}\,
         Z_1'(u,k) Z_1(u,-k) 
         + \mbox{contact terms}\,.
\end{equation}
Proceeding as in Section \ref{rcc}, we find
after comparing the expression for the correlator  with (\ref{eq:Gtxtx}) -- (\ref{eq:Gxzxz})
\begin{equation}
    G_1(\omega,q) = - \frac{\pi^2 N_c^2 T^4 {\cal B}_{(1)}(\omega,q)}
            {{\cal A}_{(1)}(\omega,q)}\,.
\end{equation}
Again, the condition ${\cal A}_{(1)}(\omega,q)=0$ is equivalent to
Dirichlet boundary condition $Z_1(u{=}0)=0$.
A typical arrangement of quasinormal frequencies
is shown in Fig.~\ref{fig:complexplane-vector}.

\begin{FIGURE}[t]
{
  \parbox[c]{\textwidth}
  {
  \begin{center}
  \psfrag{x}{$\scriptstyle \Re \wn$}
  \psfrag{y}{$\scriptstyle \Im \wn$}
  \includegraphics[width=2.8in]{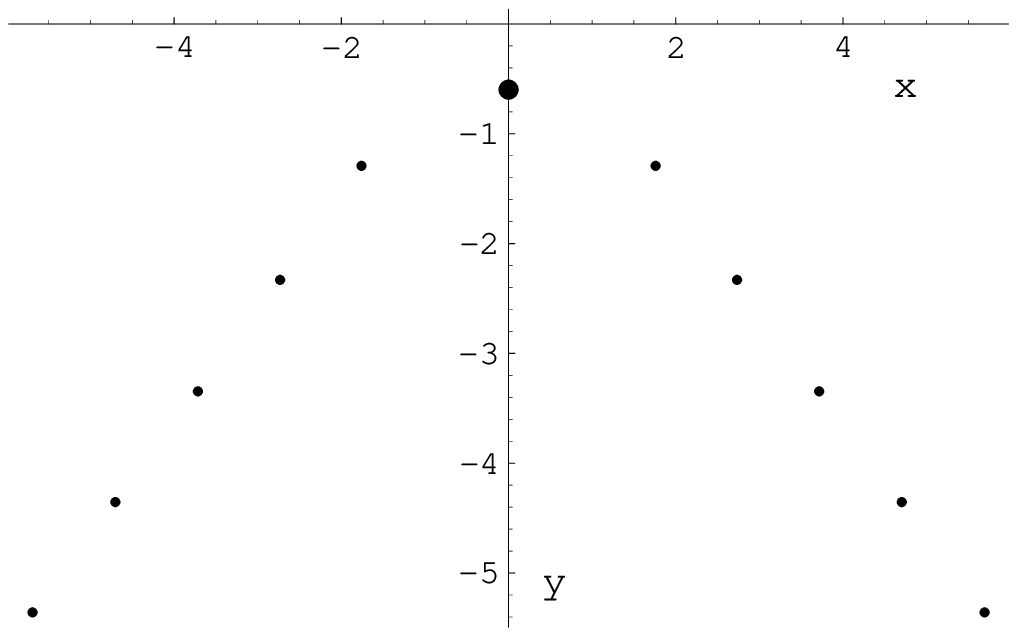}
  \caption
    {%
	Quasinormal spectrum of gravitational fluctuations
	in the shear channel, shown in the plane of complex
	$\wn\equiv\omega/2\pi T$, for spatial momentum $\qn\equiv q/2\pi T=1$.
	The quasinormal frequencies coincide with poles of
	$G_1(\omega,q)$, as explained in the text.
	As $\qn$ decreases, all poles stay at a finite distance
	away from the real axis, except for the one marked
	with a large dot. This pole is purely imaginary and approaches
	the origin in the limit $\qn\to0$.
	The presence of this special quasinormal frequency
	is a manifestation of the diffusive relaxation
	of transverse momentum density fluctuations
	in the dual ${\cal N}=4$ SYM theory.
    }
  \end{center}
  }
\label{fig:complexplane-vector}
}
\end{FIGURE}

\subsubsection*{Hydrodynamic approximation}

In the limit $\wn\ll1$, $\qn \ll 1$, the perturbative analytical solution to
Eq.~(\ref{okno}) satisfying the incoming wave boundary condition
at the horizon is%
\footnote{
	The argument of footnote \ref{fn:frequency-scaling}
	regarding the scaling
	of $\wn$, $\qn$ also applies here. The boundary condition constrains
	$\wn$ and  $\qn$ in such a way that the initial assumption that
	$\wn$ and  $\qn$ are of the same order is invalid.
	Using the correct scaling we find that
	the result (\ref{eq:shear-dispersion}) remains unchanged.
}
\begin{equation}
    Z_1 (u) = C_1 f(u)^{-{i\wn}/{2}} \left( 1 + \frac{i\qn^2 f}{2\wn} +
              O(\wn^2, \qn^2, \wn\qn)\right)\,,
\end{equation}
where $C_1$ is a normalization constant.
Expanding for small $u$, we find the connection coefficients
\begin{equation}
    {\cal A}_{(1)} = 1 + \frac{i\qn^2}{2\wn} +   O(\wn^2, \qn^2, \wn\qn)\,,
    \quad\quad
    {\cal B}_{(1)} = \frac{i(\wn^2 -\qn^2)}{2\wn}
    + O(\wn^2, \qn^2, \wn\qn)\,.
\end{equation}
The Dirichlet condition $Z_1(u{=}0)=0$ gives the
hydrodynamic quasinormal frequency
$\wn=-i\qn^2/2+O(\qn^3)$.
It is interpreted as the dispersion relation for the shear mode,
\begin{equation}
    \omega = - i \gamma_\eta q^2 + O(q^3)\,,
\label{eq:shear-dispersion}
\end{equation}
where $\gamma_\eta=1/4\pi T$.
For the function $G_1(\omega,q)$ in this approximation we find
\begin{equation}
    G_1(\omega,q) = \frac{\pi N_c^2 T^3 (\omega^2-q^2)}
                         {4 ( i \omega - q^2/4\pi T)}\,,
\label{G1}
\end{equation}
in agreement with the result obtained earlier in
\cite{Policastro:2002se}.
The quasinormal spectrum for frequencies beyond the hydrodynamic limit
was obtained in \cite{Nunez:2003eq}
using a slightly different approach.

\subsubsection{Sound channel}
\label{soundmode}

According to the discussion in Section \ref{sec:quasiholo},
equations obeyed by the components of the metric
 $H_{tt}=u h_{tt}/f(\pi T R)^2$,
 $H_{tz}=u h_{tz}/(\pi T R)^2$,  $H_{zz}=u h_{zz}/(\pi T R)^2$,
 $H_{aa}=u (h_{xx}+h_{yy})/(\pi T R)^2$
form a closed system of equations (in the radial gauge $h_{u\A}=0$).
These equations are lengthy, and we present them in
Appendix \ref{app2}.
Using the equations of motion (\ref{eom2}) -- (\ref{eom5}) one can show that
the gauge-invariant combination
\begin{equation}
     Z_2(u)\equiv 4 \wn \qn H_{tz} + 2 \wn^2 H_{zz} + H_{aa} \left[ \qn^2
           (2-f) - \wn^2\right] + 2 \qn^2 f H_{tt}
\end{equation}
obeys the following second-order differential equation:
\begin{eqnarray}
    Z_2'' &-& 
    \frac{3\wn^2 (1+u^2) + \qn^2 ( 2u^2 - 3 u^4 -3)}
         {u f (3 \wn^2 +\qn^2 (u^2-3))} \, Z_2'
    \nonumber \\
    &+& \frac{3 \wn^4 +\qn^4 ( 3-4 u^2 + u^4) +
    \qn^2 ( 4 u^5 - 4 u^3 + 4 u^2 \wn^2 - 6 \wn^2)}
    {u f^2 ( 3 \wn^2 + \qn^2 (u^2 -3))}\, Z_2
    = 0\,.
\label{Zequation}
\end{eqnarray}
In the limit $u\to 0$ this equation coincides with
Eq.~(\ref{scalar_eq}) obeyed
by a minimally coupled massless scalar,
and thus the behavior of the
solution $Z_2(u)$ near the boundary is given by
\begin{equation}
    Z_2(u) = {\cal A}_{(2)} ( 1 + \cdots) + {\cal B}_{(2)} u^2 +\cdots \,,
\label{eq:boundary-Z2}
\end{equation}
where ellipses denote higher powers of $u$.
Using the equations of motion (\ref{eom2}) -- (\ref{eom5}),
the relevant part of the on-shell boundary gravitational action
(quadratic in fluctuations) can be written as%
\footnote{
	The action involving the relevant components
	of the metric is written in Appendix \ref{app2}.
	In order to find $A(\wn,\qn,u)$,
	we form the difference $S_{B}^{(2)} - \int\!
	d\omega d q /(2\pi)^2 A(\wn,\qn,u)\, Z_2'(u,k) Z_2(u,-k)$,
	then use the equations of motion
	(\ref{eom2}) -- (\ref{eom5}) to eliminate all derivatives in the
	difference except $H_{tt}'$, and find $A(\wn,\qn,u)$ by requiring that
	the coefficient in front of $H_{tt}'$ should vanish.
}
\begin{equation}
    S_{B}^{(2)} = \lim_{u\to 0}\int\! \frac{d\omega d q}{(2\pi)^2} \,
                  A(\wn,\qn,u)\, Z_2'(u,k) Z_2(u,-k)
    + S_{CT}^{(2)}\,, \label{act_ct}
\end{equation}
where
\begin{equation}
    A(\wn,\qn,u) = \frac{3 N^2_c \pi^2 T^4 f(u)}
                        {32 u (3\wn^2 - \qn^2 (3-u^2))^2}\,,
\end{equation}
and the ``contact term'' part $S_{CT}^{(2)}$  does not contain
derivatives of the fluctuations (its boundary value is given
in Appendix \ref{app2}).
In order to compute the stress-tensor correlation functions,
we need to solve the ``wave equation''
(\ref{Zequation}) for $Z_2(u)$ subject to the boundary condition
\begin{equation}
   Z_2(u{=}0) = 4 \wn \qn H_{tz}^0 + 2 \wn^2  H_{zz}^0 +
   H_{aa}^0 \left( \qn^2 - \wn^2\right) + 2 \qn^2 H_{tt}^0\,,
\end{equation}
substitute the result into the action (\ref{act_ct}) and take the
appropriate functional derivatives%
\footnote{
	For correct normalization of the coupling
	between the boundary gravitational
	fluctuations and the stress-energy tensor,
	see Eq.~(3.18) of \cite{Policastro:2002tn} .
}
with respect to the boundary values of
the fields $H_{tt}^0$, $H_{tz}^0$, $H_{zz}^0$, $H_{aa}^0$. For
example, for the correlator $G_{tt,tt}$ we have
\begin{equation}
    G_{tt,tt} = -4\,\frac{\delta^2 S_B^{(2)}}
                {\delta H_{tt}^0(k) \delta H_{tt}^0(-k)}\,.
\end{equation}
Using the expansion (\ref{eq:boundary-Z2}),
we find that the correlators are given by
Eqs.~(\ref{eq:Gtttt}) -- (\ref{eq:Gttxx}) with%
\footnote{
	Terms analytic in $\wn$ and $\qn$ are ignored,
	even if they are divergent as $u\to0$.
}
\begin{equation}
    G_2 (\omega,q) = - \frac{N^2_c \pi^2 T^4 {\cal B}_{(2)}(\omega,q)}
                       {{\cal A}_{(2)}(\omega,q)}
    + \mbox{contact terms}\,.
\end{equation}
The problem of computing correlation functions in the dual theory
is thus reduced to finding the connection coefficients ${\cal A}_{(2)}$
and ${\cal B}_{(2)}$ of the second order ODE (\ref{Zequation}).
Zeroes of ${\cal A}_{(2)}(\wn,\qn)$ appear as poles of the correlators.
Finding the poles
is therefore equivalent to solving the boundary value problem for
the gauge invariant variable $Z_2(u)$ obeying the incoming wave
boundary condition at the horizon $u{=}1$ and Dirichlet condition
$Z_2(u{=}0)=0$ at the boundary. 

For $\qn =0$, eqs. (\ref{scalar_eq}), (\ref{okno}), (\ref{Zequation})
all reduce to the equation for a minimally coupled massless scalar
at zero spatial momentum, and
consequently all have the same quasinormal spectrum
with the asymptotics (\ref{asymptotics}).
This is expected, in accord with rotation invariance in the dual field theory,
see Eq.~(\ref{eq:small-q-stress}).
For $\qn\neq0$, the spectrum can be found numerically,
as explained in Appendix \ref{app:numerics};
a typical arrangement of quasinormal frequencies
is shown in Fig.~\ref{fig:complexplane-tensor}.
\begin{FIGURE}[t]
{
  \parbox[c]{\textwidth}
  {
  \begin{center}
  \psfrag{x}{$\scriptstyle \Re \wn$}
  \psfrag{y}{$\scriptstyle \Im \wn $}
  \includegraphics[width=2.8in]{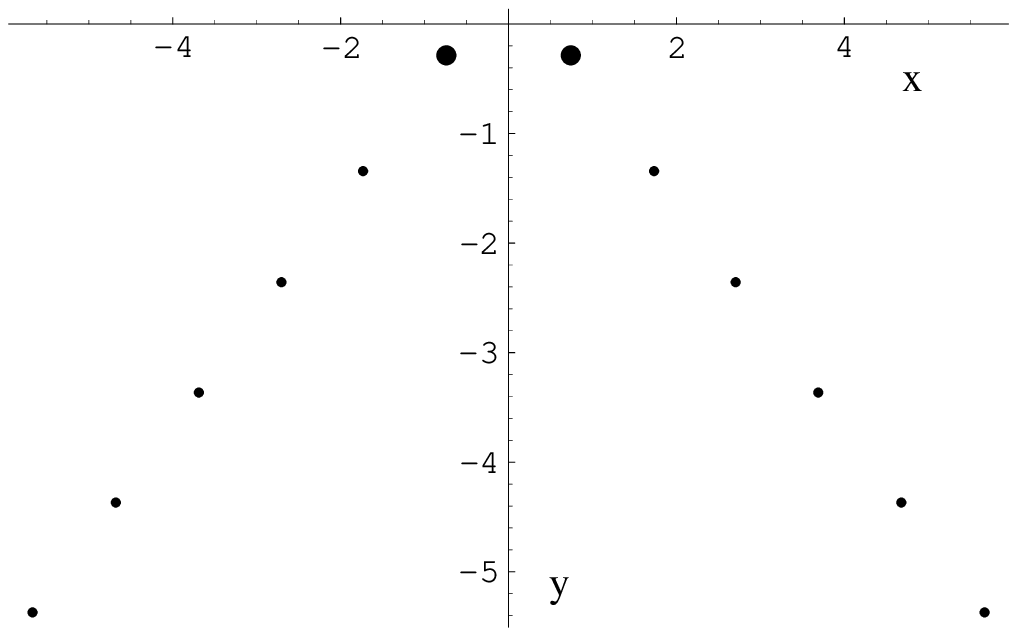}
  \caption
    {%
	Quasinormal spectrum of gravitational fluctuations
	in the sound channel, shown in the plane of complex
	$\wn\equiv\omega/2\pi T$, for spatial momentum $\qn\equiv q/2\pi T=1$.
	The quasinormal frequencies coincide with the poles of $G_2(\omega,q)$,
	as explained in the text.
	As $\qn$ decreases, all poles stay at a finite distance
	away from the real axis, except for the ones marked with
	large dots, which approach the origin as $\qn\to0$
	(see Appendix \ref{app:numerics} for the corresponding
	dispersion curves).
	Such behavior of the lowest quasinormal frequencies
	is a manifestation of oscillatory
	relaxation of longitudinal momentum density
	(as well as energy density) fluctuations in the dual
	${\cal N}=4$ SYM theory.
    }
  \end{center}
  }
\label{fig:complexplane-tensor}
}
\end{FIGURE}%
However, for small momenta, the lowest quasinormal frequency
can be found analytically.

\subsection*{Hydrodynamic approximation}

In the hydrodynamic regime $\wn \ll 1$, $\qn \ll 1$,
Eq.~(\ref{Zequation}) for $Z_2(u)$
can be solved perturbatively in $\wn$ and $\qn$.
Introducing the book-keeping parameter $\lambda$,
rescaling $\wn \rightarrow \lambda \wn$, $\qn \rightarrow \lambda \qn$,
and expanding in $\lambda \ll 1$, to first order in $\lambda$ we find
\begin{equation}
    Z_2(u) = C_2 f(u)^{-i\wn/2} \left[ \frac{\qn^2(1+u^2) - 3\wn^2}{4\qn^2}
           - \frac{i \wn f(u)}{2}\right] \,,
\label{reshz}
\end{equation}
where $C_2$ is a normalization constant. Imposing Dirichlet
boundary condition $Z_2(u{=}0)=0$ gives the lowest ($|\wn |\ll 1$)
overtone of the quasinormal spectrum
\begin{equation}
    \wn = \pm\frac{\qn}{\sqrt{3}} - \frac{i\qn^2}{3} + O\left(\qn^3\right)\,.
\label{vs}
\end{equation}
In the holographically dual finite temperature quantum field
theory, Eq.~(\ref{vs}) appears as a pole
in the retarded correlator of stress-energy tensors, and is interpreted as
dispersion relation for the sound wave mode,
\begin{equation}
    \omega (q) = \pm v_s q - i\, \Gamma_s q^2 + O(q^3)\,.
\label{eq:sound-dispersion}
\end{equation}
The values for the speed of sound $v_s=1/\sqrt{3}$ and
the attenuation constant $\Gamma_s=1/6\pi T$
coincide with those found previously in \cite{Policastro:2002tn}.
Expanding the solution (\ref{reshz}) near $u=0$ and comparing with
(\ref{eq:boundary-Z3}), we identify the coefficients
${\cal A}_{(2)}=(\qn^2 - 3\wn^2 - 2i\wn\qn^2)/4\qn^2$,
${\cal B}_{(2)} = (2\qn^2 + 5i\wn\qn^2 - 3i\wn^3)/8\qn^2$.
Thus to leading order in the hydrodynamic
approximation we obtain
\begin{equation}
    G_2 (\omega,q) = \frac{N^2_c \pi^2 T^4 q^2}{3 \omega^2 - q^2}\,.
\label{G2a}
\end{equation}
The dispersion relation $\wn = \wn(\qn)$ for the sound wave frequency
can be determined numerically (as explained in Appendix \ref{app:numerics}),
and is shown in Fig.~\ref{fig:dispersion-sound}.
\begin{FIGURE}[t]
{
\begin{minipage}[t]{.48\textwidth}
\begin{center}
  \psfrag{y}{$\scriptstyle \Re \wn $}
  \psfrag{x}{$\scriptstyle \qn$}
  \epsfig{file=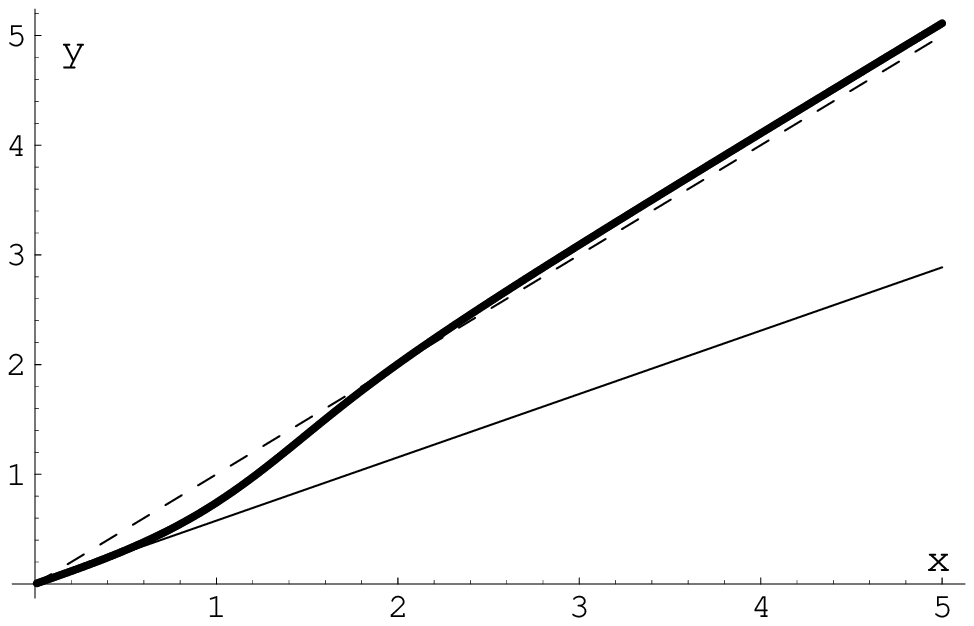,scale=0.7}
\end{center}
\end{minipage}
\begin{minipage}[t]{.48\textwidth}
\begin{center}
  \psfrag{Y}{$\scriptstyle \Im \wn$}
  \psfrag{X}{$\scriptstyle \qn$}
  \epsfig{file=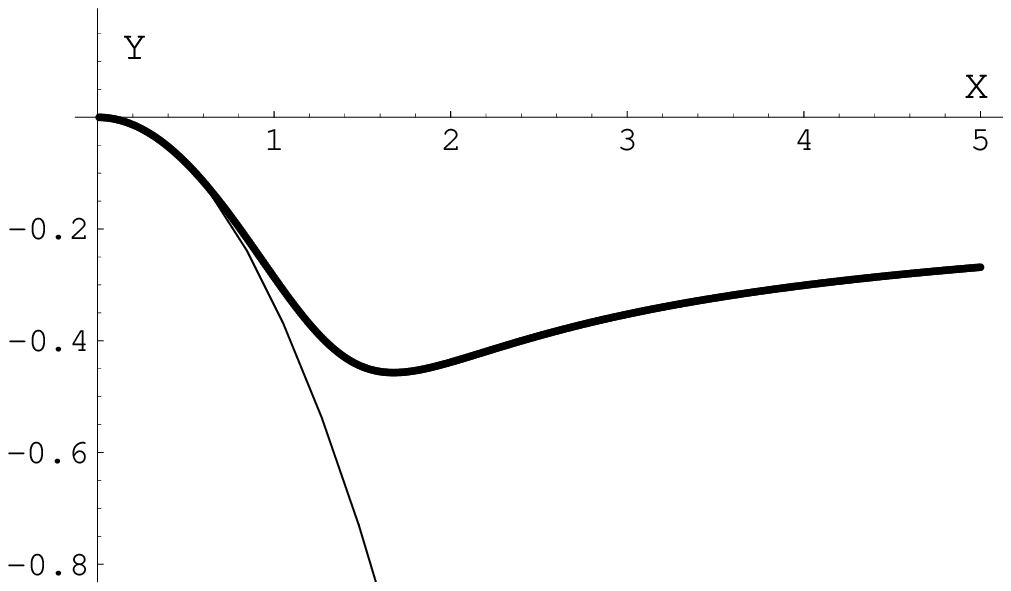,scale=0.7}
\end{center}
\end{minipage}
  \caption
    {%
     Real and imaginary parts of the lowest quasinormal (sound wave)
     frequency as a function of spatial momentum.
     Light curves correspond to
     the analytic approximation (\ref{vs}) for small $\qn$.
     Dashed line is $\wn = \qn$.
     As $\qn$ grows, the dispersion curve enters the region where
     $\partial\Re(\wn)/\partial\qn>1$.
     This, however, happens for $\qn\sim1$, where the corresponding
     singularity of the energy-momentum correlation function
     no longer has the interpretation of a sound wave.
    }
  \label{fig:dispersion-sound}
}
\end{FIGURE}

\section{Discussion}
In this paper, we proposed a general approach for identifying quasinormal
spectra of asymptotically AdS spacetimes with the poles of the retarded correlators
in the holographically dual finite temperature field theory.
Our demonstration in Section \ref{sec:applications} that
quasi-normal spectrum of gauge-invariant perturbations in
asymptotically AdS spacetimes has a precise interpretation in dual
field theory was specific to the case of five-dimensional
AdS-Schwarzschild background with a plane-symmetric event horizon.
Although we expect the same to be true in a more general setting
as discussed in Section \ref{sec:quasiholo} (such as black holes
rather than branes, or more complicated backgrounds), we did not
give a general proof.
Explicit computations for other backgrounds can be done
along the lines of Section \ref{sec:applications}.

From the dual field theory point of view,
a noteworthy observation is that the only
singularities of thermal Green's functions identified in
  the supergravity approach
are simple poles.
It may indeed be true that real-time thermal correlation functions
of gauge-invariant operators are meromorphic in complex
frequency plane; we leave the analysis of analytic structure
for future investigation.
Of course, this simple nature of singularities is possible only
at infinite $N_c$; for example, it is known that
low-energy correlation functions of conserved currents
develop branch cuts in the complex frequency plane,
with discontinuities across the cuts
suppressed in the $N_c{\to}\infty$ limit
(such cuts are not visible in classical supergravity and reflect quantum
modifications of the quasinormal spectrum) \cite{Kovtun-Yaffe}.
In a sense, the situation is similar to
zero-temperature spectrum of
confining gauge theories,
when resonances become stable as $N_c{\to}\infty$ \cite{Witten}.
However, at finite temperature, the poles of the
real-time correlation functions
can not be automatically interpreted as quasi-particles
propagating in thermal bath -- a definite interpretation
can be given only after the full spectral density
(not just the poles) is known.
In infinitely strongly coupled ${\cal N}=4$ SYM theory,
the quasiparticle interpretation is unlikely
because of the unique energy scale (temperature).
On the other hand, it
is possible that $\alpha'$ corrections to the quasinormal spectrum
will reveal new poles corresponding to heavy excitations (whose mass
scales with the 't~Hooft coupling
of the dual field theory) which can be interpreted as quasiparticles.

\begin{acknowledgments}
We would like to thank D.~T.~Son and L.~G.~Yaffe for helpful conversations,
C.~P.~Herzog for comments on the manuscript,
and J.~Mas for correspondence and comments on the paper
\cite{Policastro:2002tn}. A.~O.~S. would like to thank 
the organizers of the ``QCD and String Theory'' workshop
at the KITP, UC Santa Barbara, where part of this work was completed.
The work of P.~K.~K.  was supported in part
by the National Science Foundation under Grant No. PHY99-07949.
Research at Perimeter Institute is supported in part by funds
from NSERC of Canada.
\end{acknowledgments}

\newpage
\appendix

\section{Equations of motion and boundary action for the sound channel}
\label{app2}

Fluctuations of the sound wave mode
satisfy a system of differential equations
\begin{eqnarray}
H_{tz}' &=& \frac{2\wn \qn}{f+2} \left( H_{aa} - H_{tt}\right)
           +\frac{(u^3-3u-2\wn^2)}{\qn f (f+2)}
            \left(\wn\,H_{aa}+\wn\,H_{zz}+2\qn\,H_{tz}\right) \nonumber \\
        &-& \frac{3\wn f}{\qn  (f+2)} \, H_{tt}'\,,\label{eom2} \\
H_{aa}' &=& \frac{\wn (u^3-3 u - 2\wn^2)}{\qn^2 f^2 (f+2)} \left( \wn H_{aa} 
           +\wn H_{zz} + 2 \qn H_{tz}\right) 
           -\frac{3\wn^2 - \qn^2(f+2)}{\qn^2 (f+2)} H_{tt}' \nonumber \\ 
        &+& \frac{2\wn^2}{f(f+2)} H_{aa}
           +\frac{u^3-3u-2\wn^2}{f(f+2)}\, H_{tt}\,,\label{eom3} \\
H_{zz}' &=& \frac{2}{f (f+2)} \left[ \wn^2 H_{aa} +\wn^2 H_{zz}
           +\qn^2 f \left( H_{tt}-H_{aa} \right) + 2 \wn\qn H_{tz}\right]
           +\frac{3 f H_{tt}'}{f+2} \nonumber \\
        &+& \frac{1}{f(f+2)}\left[2\wn^2 H_{aa} + (u^3-3u-2\wn^2)H_{tt}\right]
           -\frac{3\wn^2 -\qn^2 (f+2)}{\qn^2 (f+2)} H_{tt}' \nonumber \\
        &+& \frac{\wn (u^3-3u-2\wn^2)}{\qn^2 f^2 (f+2)}
            \left(\wn H_{aa} + \wn H_{zz} + 2\qn H_{tz}\right)\,,\label{eom4}\\
H_{tt}''&=& \frac{1}{2 u^2 f^2(f+2)} \Biggl\{ 6 u f (1+u^2) H_{tt}'
           +2 u f \left[ 2 \wn^2 + \qn^2 (1+u^2)\right] H_{aa} 
           +4\wn^2 u f H_{zz} \nonumber \\ 
        &+& 8 \wn \qn u f H_{tz} + 4 u f^2 \qn^2 H_{tt}\Biggr\}\,.\label{eom5}
\end{eqnarray}
The part of the boundary action quadratic in fluctuations is
\cite{Policastro:2002tn}
\begin{align}
\label{action_onshell}
S_{B}^{(2)} = \lim_{u\to0}
              & \frac{\pi^2 N_c^2 T^4}{8} \int\! d^4x\,  \Biggl[ 
                \frac18 \left( 3 H_{tt}^2  - 12 H_{tz}^2 + 2 H_{tt}H_{ii}
                + 2 H_{zz}H_{aa} - H_{zz}^2 \right) \nonumber \\
              & - \frac{f(u)}{2 u}\left( H_{tz}^2 + \frac14 H_{aa}^2
                -  H_{tt}H_{ii} + H_{zz}H_{aa}\right)' \Biggr] \,,
\end{align}
where prime denotes the derivative with respect to $u$,
and expressions such as $H_{tt}^2$ are to be understood as
$H_{tt}(u,k) H_{tt}(u,-k)$.
The boundary value of the ``contact term'' part of the gravitational action
(\ref{act_ct})
is given by
\begin{eqnarray}
\label{ctterms}
  S_{CT}^{(2)} (0)
  &=&  \lim_{u\rightarrow 0}  S_{CT}^{(2)} =
      -\frac{N^2_c \pi^2 T^4}{48} \Biggl[ (H_{aa}^{0})^2
      -\frac{\qn^2+3\wn^2}{2 (\qn^2-\wn^2)} H_{aa}^{0}H_{tt}^{0}
      -\frac{4\wn\qn}{\qn^2-\wn^2} H_{aa}^{0}H_{tz}^{0} \nonumber \\
  &-&  \frac{29\qn^4 - 30\wn^2\qn^2 + 9\wn^4)}{4(\qn^2-\wn^2)^2}(H_{tt}^{0})^2
      -\frac{4\wn\qn(5\qn^2- 3\wn^2)}{(\qn^2-\wn^2)^2} H_{tt}^{0}H_{tz}^{0}
       \nonumber \\
  &-&  \frac{3\qn^2+\wn^2}{2 (\qn^2-\wn^2)} H_{aa}^{0}H_{zz}^{0}
      -\frac{9 \qn^4 + 2 \wn^2 \qn^2 - 3 \wn^4}{2 (\qn^2-\wn^2)^2}
       H_{tt}^{0}H_{zz}^{0}
      -\frac{4\wn\qn (3\qn^2 - \wn^2)}{(\qn^2-\wn^2)^2} H_{tz}^{0}H_{zz}^{0}
       \nonumber \\
  &+&  \frac{3\qn^4 - 18\wn^2\qn^2 + 7\wn^4}{4(\qn^2-\wn^2)^2}(H_{aa}^{0})^2
      -\frac{3\qn^4 + 14\wn^2\qn^2 - 9\wn^4}{(\qn^2-\wn^2)^2} (H_{tz}^{0})^2
       \Biggr] \,.
\end{eqnarray}
Equation (\ref{ctterms}) appears to contain more than just contact
terms, since there is a pole at $|\wn|=\qn$. This pole, however,
is artificial --- it reflects the normalization of $Z_2$, and
cancels in the final expression for the correlators.

If one chooses to keep track of contact terms
in field theory correlators, one can use
$S_{CT}^{(2)}$ to reproduce contact terms
computed earlier in \cite{Policastro:2002tn}.
For example,
\begin{equation}
   G_{tt,tt} = \frac{2 q^4}{3 (q^2 - \omega^2)^2} G_2(\omega,q)
   -4 \,\frac{\delta^2 S_{CT}^{(2)}(0)}{(\delta  H_{tt}^0)^2}\,
\label{cortttt}
\end{equation}
 in the hydrodynamic limit becomes
\begin{equation}
   G_{tt,tt} = \frac{3 N_c^2 \pi^2 T^4 (3\omega^2-5 q^2)}
                    {8 (q^2 - 3 \omega^2)}\,,
\end{equation}
which coincides with the result of \cite{Policastro:2002tn}.

\section{Frobenius solution}
\label{app:numerics}

To find the full quasinormal spectrum, one has to analyze
wave equations (\ref{eq:electric-fields}),
(\ref{scalar_eq}), (\ref{okno}), and (\ref{Zequation}) in more detail. 
All these equations are Fuchsian ODEs with $k$ singular points,%
\footnote{
	For the transverse Maxwell equation (\ref{eqE_x})
	satisfied by $E_\alpha$,
	and for the scalar channel wave equation (\ref{scalar_eq})
	satisfied by $Z_3$,
	the number of singular points is $k{=}4$,
	corresponding to $u=0,\pm1,\infty$.
	For the longitudinal Maxwell equation (\ref{eqE_z}) satisfied by $E_z$,
	and the shear channel wave equation (\ref{okno}) satisfied by $Z_1$,
	the number of singular points is $k{=}6$, corresponding to
	$u=0,\pm1,\pm\sqrt{1-\wn^2/\qn^2},\infty$.
	For the sound channel wave equation (\ref{Zequation})
	satisfied by $Z_3$,
	the number of singular points is also $k{=}6$,
	corresponding to $u=0,\pm1,\pm\sqrt{3(1-\wn^2/\qn^2)},\infty$.
}
two of which correspond, respectively,
to the horizon ($u=1$) and the boundary ($u=0$).
For all of the above equations,
the exponents at the horizon are equal to $\pm i\wn/2$, corresponding to two
local solutions representing outgoing (incoming) waves.
The solution obeying incoming wave boundary condition
at the horizon can be represented
as a power series around $u{=}1$,
\begin{equation}
    Z(u) = (1-u)^{-{i\wn}/{2}} (1+u)^{-{\wn}/{2}}
             \sum_{n=0}^{\infty} a_n (\wn,\qn) (1-u)^n\,,
\label{eq:Frobenius-expansion}
\end{equation}
where $Z(u)$ stands for either of
$E_\alpha$, $E_z$, $Z_1$, $Z_2$, $Z_3$.
The coefficients $a_n$ of the series expansion
obey $(k{-}1)$-term recursion relations 
which can be found by substituting (\ref{eq:Frobenius-expansion})
in the original differential equations.
Quasinormal spectrum is determined by imposing
Dirichlet boundary condition at $u=0$,
\begin{equation}
    Z(0) =  \sum_{n=0}^{\infty} a_n (\wn,\qn)  = 0\, \label{num}
\end{equation}
and solving Eq.~(\ref{num}) numerically
taking a sufficiently large
but finite number of terms in the sum.%
\footnote{
Careful readers may note that the expansion
(\ref{eq:Frobenius-expansion})
is guaranteed to converge only inside a circle of radius $\rho$
around $u=1$ (in the complex $u$ plane), where $\rho$
is the distance to the nearest singular point, which may become
less than one (distance to the boundary)
for some values of $\wn$ and $\qn$.
However, even for such values of $\wn$ and $\qn$,
our numerical results for diffusive and shear modes
are in agreement with previous calculations
\cite{Nunez:2003eq} where this issue did not arise,
thus suggesting wider applicability of the expansion
(\ref{eq:Frobenius-expansion}).
}
The spectra of all of the above wave equations
are qualitatively similar, except for the
behavior of the lowest (hydrodynamic) frequency
which is absent for $E_\alpha$ and $Z_3$.
For $E_z$ and $Z_1$, hydrodynamic frequencies
are purely imaginary (given by Eqs.
(\ref{eq:diffusion-dispersion}) and (\ref{eq:shear-dispersion})
for small $\omega$ and $q$),
and presumably move off to infinity as $q$ becomes large.
For $Z_2$, the hydrodynamic frequency has
both real and imaginary parts (given by
Eq.~(\ref{eq:sound-dispersion}) for small $\omega$ and $q$),
and eventually (for large $q$) becomes indistinguishable
in the tower of other eigenfrequencies.
As an example, dispersion relations for the
three lowest quasinormal frequencies in the sound channel
(including the one of the sound wave)
are shown in Fig.~\ref{fig:dispersion-obertones}.
\begin{FIGURE}[t]
{
\begin{minipage}[t]{.48\textwidth}
\begin{center}
  \psfrag{X}{$\scriptstyle \qn$}
  \psfrag{Y}{$\scriptstyle \Re\wn$}
  \epsfig{file=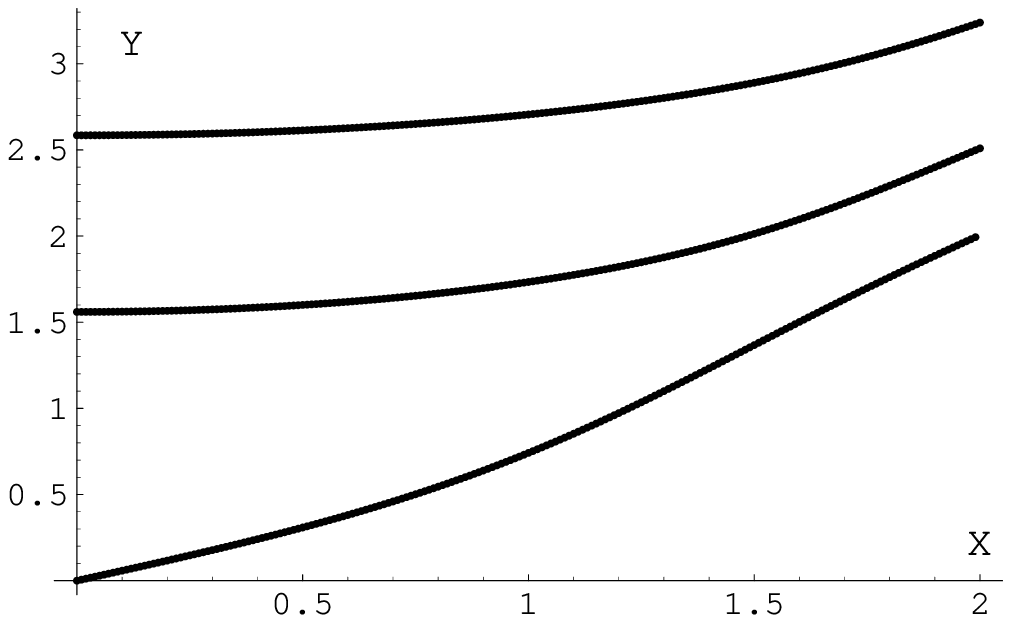,scale=0.7}
\end{center}
\end{minipage}
\begin{minipage}[t]{.48\textwidth}
\begin{center}
  \psfrag{X}{$\scriptstyle \qn$}
  \psfrag{Y}{$\scriptstyle \Im \wn$}
  \epsfig{file=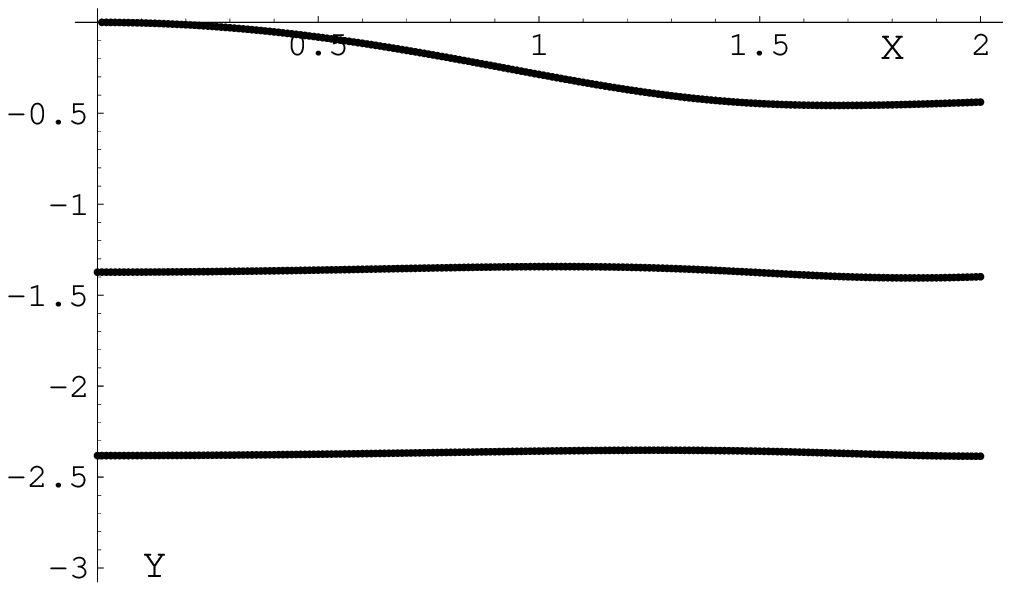,scale=0.7}
\end{center}
\end{minipage}
  \caption
    {%
	Real and imaginary parts of three lowest quasinormal
	frequencies as function of spatial momentum.
	The curves for which $\wn{\to}0$ as $\qn{\to}0$
	correspond to hydrodynamic sound mode in the dual
	finite temperature ${\cal N}{=}4$ SYM theory.}
  \label{fig:dispersion-obertones}
}
\end{FIGURE}%
The tables below give numerical values of
quasinormal frequencies for $\qn=1$. 
Only non-hydrodynamic frequencies are shown in the tables.
The position of hydrodynamic frequencies at $\qn=1$ is
$\wn=-3.250637 i$ for the $R$-charge diffusive mode,
$\wn = -0.598066 i $ for the shear mode, and
$\wn = \pm 0.741420 - 0.286280 i$ for the sound mode.
The numerical values of the lowest five
(non-hydrodynamic) quasinormal frequencies
for electromagnetic perturbations are:
\begin{center}
\begin{tabular}{|c||c|c||c|c|}
\hline
{} &
\multicolumn{2}{c||}{Transverse channel} &
\multicolumn{2}{c|}{Diffusive channel} \\
\cline{2-5}
$n$ & $\Re\wn$ & $\Im\wn$ & $\Re\wn$ & $\Im\wn$ \\
\hline
$1$ & $\pm$1.547187 & $-$0.849723 & $\pm$1.147831 & $-$0.559204 \\ 
$2$ & $\pm$2.398903 & $-$1.874343 & $\pm$1.910006 & $-$1.758065 \\
$3$ & $\pm$3.323229 & $-$2.894901 & $\pm$2.903293 & $-$2.891681 \\
$4$ & $\pm$4.276431 & $-$3.909583 & $\pm$3.928555 & $-$3.943386 \\
$5$ & $\pm$5.244062 & $-$4.920336 & $\pm$4.946818 & $-$4.965186 \\
\hline
\end{tabular}
\end{center}
%The numerical values of the lowest five
%(non-hydrodynamic) quasinormal frequencies
and for gravitational perturbations are:
\begin{center}
\begin{tabular}{|c||c|c||c|c||c|c|}
\hline
{}  & 
\multicolumn{2}{c||}{Scalar channel} & 
\multicolumn{2}{c||}{Shear channel} & 
\multicolumn{2}{c|}{Sound channel} \\ \cline{2-7}
$n$ & $\Re\wn$ & $\Im\wn$ & $\Re\wn$ & $\Im\wn$ & $\Re\wn$ & $\Im\wn$\\
\hline
%$1$ & $\pm$1.672019 & $-$1.340413 & $\pm$1.759116 & $-$1.291594 & $\pm$1.733511 & $-$1.343008 \\
%$2$ & $\pm$2.665855 & $-$2.358919 & $\pm$2.733081 & $-$2.330405 & $\pm$2.705540 & $-$2.357062 \\
%$3$ & $\pm$3.659438 & $-$3.366590 & $\pm$3.715933 & $-$3.345343 & $\pm$3.689392 & $-$3.363863 \\
%$4$ & $\pm$4.654341 & $-$4.370866 & $\pm$4.703643 & $-$4.353487 & $\pm$4.678736 & $-$4.367981 \\
%$5$ & $\pm$5.650322 & $-$5.373631 & $\pm$5.694472 & $-$5.358205 & $\pm$5.671091 & $-$5.370784 \\
$1$ & $\pm$1.954331 & $-$1.267327 & $\pm$1.759116 & $-$1.291594 & $\pm$1.733511 & $-$1.343008 \\
$2$ & $\pm$2.880263 & $-$2.297957 & $\pm$2.733081 & $-$2.330405 & $\pm$2.705540 & $-$2.357062 \\
$3$ & $\pm$3.836632 & $-$3.314907 & $\pm$3.715933 & $-$3.345343 & $\pm$3.689392 & $-$3.363863 \\
$4$ & $\pm$4.807392 & $-$4.325871 & $\pm$4.703643 & $-$4.353487 & $\pm$4.678736 & $-$4.367981 \\
$5$ & $\pm$5.786182 & $-$5.333622 & $\pm$5.694472 & $-$5.358205 & $\pm$5.671091 & $-$5.370784 \\
\hline
\end{tabular}
\end{center}

\end{document}